\documentclass[journal=jacsat,manuscript=article,layout=traditional]{achemso}
\setkeys{acs}{usetitle=true}

\usepackage[version=3]{mhchem} 

\usepackage[colorlinks=false, linktocpage=true]{hyperref} 

\usepackage{xcolor,lipsum}
\usepackage{changepage}
\usepackage{verbatim}
\usepackage{makecell}
\definecolor{darkviolet}{rgb}{0.58, 0.0, 0.83}
\definecolor{darkpurple}{rgb}{0.4, 0.0, 0.4}
\definecolor{applegreen}{rgb}{0.55, 0.71, 0.0}
\definecolor{ferrarired}{rgb}{1.0, 0.11, 0.0}
\definecolor{darkorange}{rgb}{1.0, 0.55, 0.0}
\definecolor{cadmiumyellow}{rgb}{1.0, 0.96, 0.0}
\definecolor{electricyellow}{rgb}{1.0, 1.0, 0.0}
\definecolor{fluorescentyellow}{rgb}{0.8, 1.0, 0.0}
\definecolor{frenchblue}{rgb}{0.0, 0.45, 0.73}
\definecolor{lightgray}{gray}{0.8}
\definecolor{maroon}{cmyk}{0,0.87,0.68,0.32}

\newcommand{\myhcolor}[1]{\textbf{\textcolor{black}{#1}}}
\newcommand{\RNum}[1]{\uppercase\expandafter{\romannumeral #1\relax}}
\usepackage{tikz}
\usepackage{titlesec}
\definecolor{mydarkblue}{rgb}{0.00, 0.00, 0.60}
\definecolor{mygrey}{gray}{.8}

\titleformat{\section}
  {\color{mydarkblue}\normalfont\Large\bfseries} 
  {}                                             
  {0pt}                                          
  {\thesection.\quad}
\titleformat{\subsection}
  {\color{mydarkblue}\normalfont\large\bfseries}
  {}
  {0pt}
  {\thesubsection\quad}

\titleformat{name=\section,numberless}
  {\normalfont\bfseries}
  {}
  {0pt}
  {}
\titleformat{name=\subsection,numberless}
  {\normalfont\bfseries}
  {}
  {0pt}
  {}

\author{Mike Devereux}
\affiliation{Department of Chemistry, University of Basel, Klingelbergstrasse 80, CH-4056 Basel, Switzerland} 
\author{Marco Pezzella}
\affiliation{Department of Chemistry, University of Basel, Klingelbergstrasse 80, CH-4056 Basel, Switzerland} 
\author{Shampa Raghunathan}
\affiliation{Department of Chemistry, University of Basel, Klingelbergstrasse 80, CH-4056 Basel, Switzerland}  
\author{Markus Meuwly}
\affiliation{Department of Chemistry, University of Basel, Klingelbergstrasse 80, CH-4056 Basel, Switzerland}  
\alsoaffiliation{Department of Chemistry, Brown University, Providence, RI, USA}
\email{m.meuwly@unibas.ch}

\title{Polarizable Multipolar Molecular Dynamics Using Distributed Point
  Charges}

\abbreviations{MD, TIP}

\begin{document}

\begin{abstract}
Distributed point charge models (DCM) and their minimal variants
(MDCM) have been integrated with tools widely used for condensed-phase
simulations, including a virial-based barostat and a slow-growth
algorithm for thermodynamic integration. Minimal DCM is further
developed with a systematic approach to reduce fitting errors in the
electrostatic interaction energy and a new fragment-based approach
offers considerable speedup of the MDCM fitting process for larger
molecules with increased numbers of off-centered charged
sites. Finally, polarizable (M)DCM is also introduced in the present
work. The developments are used in condensed-phase simulations of
popular force fields with commonly applied simulation
conditions. (M)DCM equivalents for a range of widely used water force
fields and for fluorobenzene (PhF) are developed and applied along
with the original models to evaluate the impact of reformulating the
electrostatic term. Comparisons of the molecular electrostatic
potential (MEP), electrostatic interaction energies, and bulk
properties from molecular dynamics simulations for a range of models
from simple TIP$n$P ($n = 3$--5) to the polarizable, multipolar
iAMOEBA models for water and an existing quadrupolar model for PhF
confirm that DCMs retain the accuracy of the original models,
providing a homogeneous, efficient, and generic point charge
alternative to a multipolar electrostatic model for force field
development and multilevel simulations.
\end{abstract}

\section{INTRODUCTION}
Empirical force fields (FFs) are routinely used for simulating a
multitude of chemical and biochemical phenomena\cite{case2005amber,
  jorgensen2005molecular,van2005gromacs,christen2005gromos,charmminecc}.
Commonly employed FFs divide interactions into intra- and
intermolecular terms that include point charges (PCs) for Coulomb
interactions. While interactions between nuclear-centered point
charges are rapid to evaluate which allows application to large
condensed-phase systems and relatively long timescales, their accuracy
is compromised as they do not correctly describe charge density
anisotropy\cite{stonebook}. A new generation of FFs aims to overcome
these drawbacks, either using higher-order multipolar
electrostatics\cite{MM08mtp,cardamone2014multipolar} in methods such
as AMOEBA (atomic multipole optimized energetics for biomolecular
applications)\cite{ren2003polarizable,ponder2010current,ren2011polarizable,laury2015revised},
SIBFA (sum of interaction between fragments ab initio
computed)\cite{gresh2007anisotropic,SIBFA1,SIBFA2,SIBFA3,SIBFA4} and
QCTFF (Quantum Chemical Topological Force Field)\cite{QUA:QUA24900},
or using Gaussian functions to directly describe the underlying charge
density in methods such as EFP (effective fragment
potential)\cite{gordon2012fragmentation}, GEM (gaussian electrostatic
model)\cite{cisneros2012application,duke2014gem}, and NEMO
(nonempirical molecular orbital)\cite{engkvist2000accurate}. On the
other hand, the use of higher-order atomic multipoles, while resulting
in improved accuracy, introduces non-negligible computational overhead
due to the additional complexity and increased number of terms that
need to be
evaluated\cite{handley2009optimal,MM.mtp:2012,kramer2013multipole,devereux2014novel}.\\

\noindent
An alternate tractable approach is to represent the MEP as a truncated
multipole expansion transformed into a set of appropriately
distributed point charges. Charges can be placed in fixed arrangements
relative to the nuclei,\cite{devereux2014novel} or machine learning
can be used to replace fixed arrangements with a minimal set of
optimally positioned off-center charges.\cite{mdcm} Recently it was
demonstrated that these Distributed Charge Models
(DCMs)\cite{devereux2014novel} and Minimal Distributed Charge Models
(MDCMs)\cite{mdcm} can be implemented into widely used molecular
dynamics software as an alternative to conventional, PC-based energy
terms. The use of point charges for representing charge density
anisotropy reduces the complexity of Coulomb terms relative to a
traditional multipolar formalism considerably, allowing efficient MD
simulations while maintaining the accuracy of a truncated multipole
expansion.\\

\noindent
The versatility of (M)DCMs (i.e. distributed charge models with and
without machine learning optimization) additionally yields a
homogeneous implementation of different types of electrostatic models
(nuclear-centered charges, off-center charges and multipolar
electrostatics) using a single routine, with combinations of models of
different complexities in a single simulation -- so-called
`multilevel' simulations. Its compatibility with remaining standard
bonded and non-bonded FF terms promises to make adaptation of
next-generation (multipolar) FF electrostatics straightforward in
widely used simulation
packages.\cite{charmm,case2005amber,van2005gromacs,openMM}\\

\noindent
The ability to generate models of increasing accuracy by increasing
the number of charges in an MDCM fit offers an important tool to force
field developers to carefully balance the accuracy of a model with the
computational cost incurred from adding each additional charge. Models
for the moiety or moieties of interest, such as a solute or protein
ligand and immediate environment, can be created at the highest level
of detail, while remaining interacting species can be optimally fitted
to balance accuracy in the potential energy surface with computational
efficiency to reach the system sizes and timescales required to
sufficiently sample the relevant phase space. This multilevel approach
is akin to the more familiar mixed quantum mechanical/molecular
mechanical (QM/MM) treatments which also employ methods of different
accuracy for different parts of a simulation
system.\cite{merz2014QMMM} Similarly, promising new models can equally
be incorporated from one force field into another by transferring
their parameters, and by refitting parameters of interacting neighbors
at a level of detail that works optimally with that model, rather than
combining existing models of different complexities that may not be
compatible.\\

\noindent
In this work recent advances in the implementation of (M)DCM are
exploited and applied to condensed-phase simulations, including an
intermediate fragment fitting step to improve efficiency of the MDCM
fitting process for larger systems, improved error handling based on
analysis of the relationship between errors in the fitted MEP and
errors in the electrostatic interaction energy, integration with
barostats for simulations in the isothermal-isobaric ensemble and
slow-growth routines for thermodynamic integration
calculations. Different (M)DCM models are developed to replace the
electrostatic terms in several water force fields commonly used in
chemical and biomolecular simulations and in a multipolar force field
for PhF. In the first section of the results (M)DCM representations
are generated for water models of increasing complexity ranging from
the simple but widely used ``TIP3P"\cite{jorgensen1983comparison} to
the multipolar, polarizable ``iAMOEBA"
potential\cite{wang2013systematic}. All terms of each original force
field other than the electrostatic term are left untouched, requiring
particularly close agreement with the original charge model to avoid
reparametrization. Then, a similar approach is taken for the PhF
molecule, demonstrating the applicability to a solute molecule in a
condensed-phase aqueous environment. Comparisons of energies and bulk
properties in each case are used to demonstrate the accuracy of a
distributed charge approach with respect to a more computationally complex
multipolar description of molecular electrostatics.\\

\section{Background}
\subsection{DCM}
Multipolar force fields are based on the fact that any charge
distribution can be represented as a series expansion, where the
successive terms are multipole moments of increasing
rank.\cite{stonebook,Price2004MTPFF} Nuclear centers are typically
used as convenient origins to locate `atomic multipoles', as is the
case in the ``distributed multipole analysis''
(DMA)\cite{stone2005GDMA,stone1985DMA}, ``Atoms in Molecules''
(AIM)\cite{cardamone2014multipolar}, and
AMOEBA,\cite{ponder2010current} to improve convergence of the
multipole expansion as opposed to using a single molecular origin. As
the rank of these terms increases from dipole to quadrupole to
octupole and beyond, the contribution that they make to the electric
field in regions beyond the extent of the original charge distribution
decays increasingly rapidly with distance. It is therefore often
possible to truncate the series expansion at the atomic quadrupole
moment\cite{spackman1986mtpTruncation,misquitta2014stockholder} while
maintaining accuracy of the electrostatic potential outside the
molecular surface. This is especially true if the terms of the
expansion are fitted to the electrostatic potential rather than
derived directly from the electron
density.\cite{kramer2013multipole,sibfaFit}\\

\noindent
For a discrete distribution of $n$ charges a multipole expansion
truncated at the quadrupole moment can be expressed using spherical
harmonics as:
\begin{eqnarray}
\begin{aligned}[c]
Q_{00}&=\sum_{i=1}^{n} q_{i} \\ Q_{10}&=\sum_{i=1}^{n} q_{i} r_{z,i}
\\ Q_{11c}&=\sum_{i=1}^{n} q_{i} r_{x,i} \\
\end{aligned}
\hspace{2 cm}
\begin{aligned}[c]
Q_{11s}&=\sum_{i=1}^{n} q_{i} r_{y,i} \\
Q_{20}&=\sum_{i=1}^{n}  \frac{1}{2}q_{i}(3 r_{z,i}^{2}-r^{2})  \\
Q_{22c}&=\sum_{i=1}^{n} \sqrt{\frac{3}{4}} q_{i}(r_{x,i}^{2}-r_{y,i}^{2}) 
\end{aligned}
\label{Eq:q-simul}
\end{eqnarray}
where $q_{i}$ is point charge $i$, $r_{x,i}$ is the $x$-coordinate of
$i$ and $Q_{lm}$ is the total atomic multipole moment of rank
$(l,m)$.\cite{stonebook} For a continuous charge density an analogous
volume integral replaces the summation.\\

\noindent
DCM is based on the fact that the converse is also true, i.e. any
truncated multipole expansion, even one derived from a continuous
charge density, can be represented by a suitable arrangement of
discrete point charges.\cite{devereux2014novel,Gao2014165} An
illustrative but general example is an octahedral charge arrangement,
where the magnitude of the charge at each vertex of the octahedron is
analytically determined to exactly reproduce all multipole moments up
to quadrupole according to:
\begin{eqnarray}
\begin{aligned}[c]
q_{(d_{\rm q},0,0)}&=\frac{Q_{00}}{6}+\frac{Q_{11c}}{2d_{\rm q}}-
\frac{Q_{20}}{6d_{\rm q}^{2}}+\frac{Q_{22c}}{2\sqrt{3}d_{\rm
    q}^{2}}\\ q_{(-d_{\rm
    q},0,0)}&=\frac{Q_{00}}{6}-\frac{Q_{11c}}{2d_{\rm q}}-
\frac{Q_{20}}{6d_{\rm q}^{2}}+\frac{Q_{22c}}{2\sqrt{3}d_{\rm
    q}^{2}}\\ q_{(0,d_{\rm
    q},0)}&=\frac{Q_{00}}{6}+\frac{Q_{11s}}{2d_{\rm q}}-
\frac{Q_{20}}{6d_{\rm q}^{2}}-\frac{Q_{22c}}{2\sqrt{3}d_{\rm q}^{2}}\\
\end{aligned}
\hspace{2 cm}
\begin{aligned}[c]
q_{(0,-d_{\rm q},0)}&=\frac{Q_{00}}{6}-\frac{Q_{11s}}{2d_{\rm q}}-
 \frac{Q_{20}}{6d_{\rm q}^{2}}-\frac{Q_{22c}}{2\sqrt{3}d_{\rm q}^{2}}\\
 q_{(0,0,d_{\rm q})}&=\frac{Q_{00}}{6}+\frac{Q_{10}}{2d_{\rm q}}+
 \frac{Q_{20}}{3d_{\rm q}^{2}} \\
 q_{(0,0,-d_{\rm q})}&=\frac{Q_{00}}{6}-\frac{Q_{10}}{2d_{\rm q}}+
 \frac{Q_{20}}{3d_{\rm q}^{2}}
\end{aligned}\label{Eq:q-octa}
\end{eqnarray}
Here, $d_{\rm q}$ is the fixed distance of the charges from the
nuclear coordinate of an atom. Note that the $Q_{21c}$, $Q_{21s}$ and
$Q_{22s}$ quadrupole moment components vanish if the correct local
axis system for an atom is chosen.\cite{devereux2014novel}\\

\noindent
It is therefore possible to replace all 6 nonzero multipole moments by
6 point charges. The total multipole moments of the charge
distribution will exactly match the multipole expansion up to
truncation rank, so the two will differ only by their higher order
terms, i.e. octupole and beyond. These terms can either be kept small
by reducing the distance $d_{\rm q}$, or deliberately enhanced to
potentially provide accuracy beyond the truncated multipole expansion
by fixing $d_{\rm q}$ to reproduce some of the higher order multipole
moments of the reference atom. The main advantage of such an approach
is a considerable reduction in complexity of the terms, as shown
explicitly for the quadrupole--quadrupole interaction in section 1 of
the SI. Thus, in DCM no fitting is required. Rather, the multipole
moments are converted into a distributed charge arrangement based on
analytical formulae.\\

\begin{figure}[!ht]
\includegraphics[width=8.0cm]{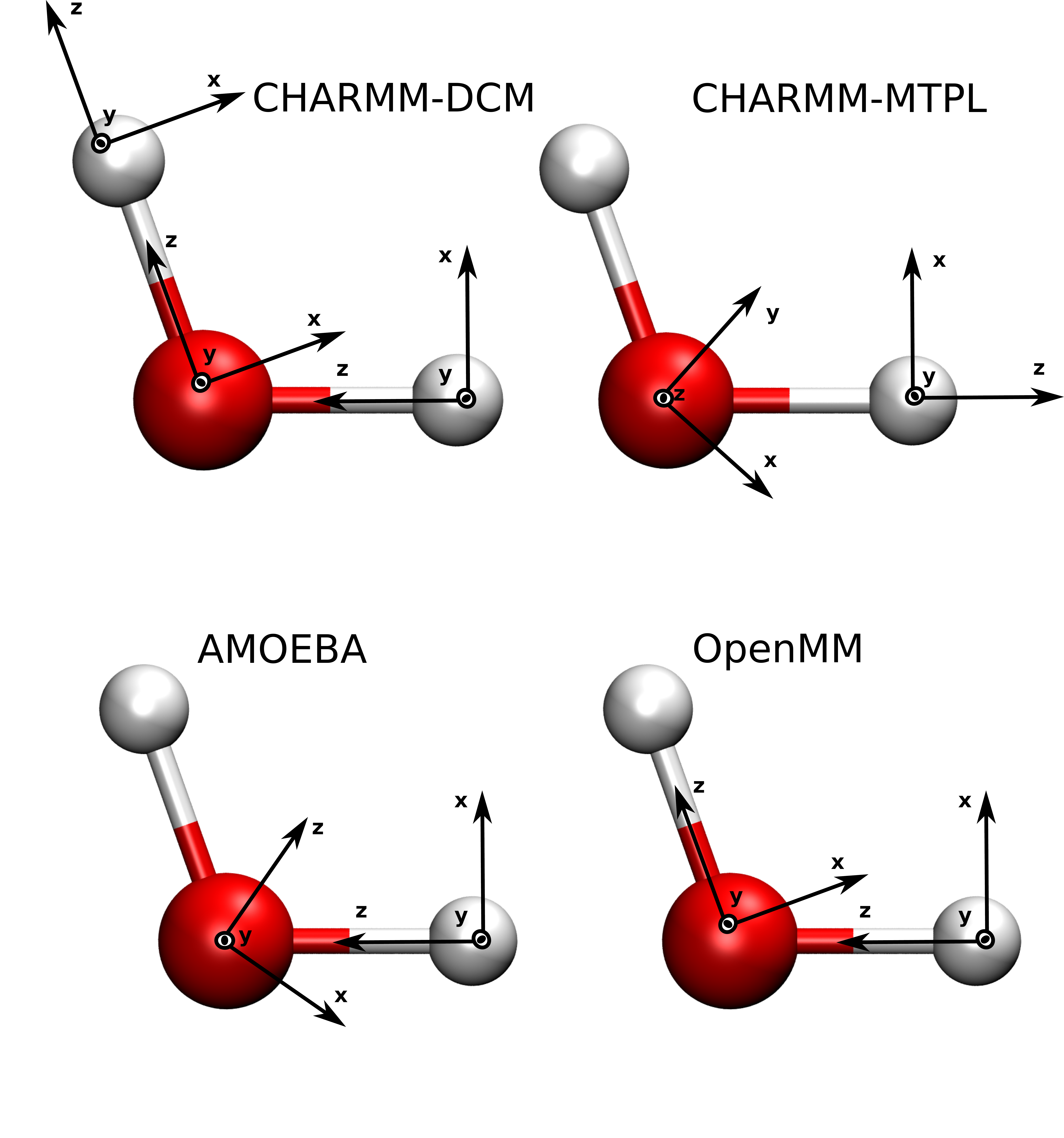}
\caption{Local axis systems required to convert between the various
  electrostatic models presented in the text and to implement them
  into CHARMM and OpenMM. H-atom axes are equivalent except for the
  case of DCM, where one H-atom shares the same local axis system as
  its bonded O-atom neighbor.}
\label{fig:lra}
\end{figure}

\noindent
For both multipolar and DCM approaches torques are generated by the
off-centered charges or multipole moments that need to be distributed
across the surrounding nuclei. As is generally the case in multipolar
force fields, the torques of DCM models are applied to the nuclei that
define the local (atomic) axis system of each charge (Figure
\ref{fig:lra}), as described elsewhere.\cite{devereux2014novel}\\

\subsection{MDCM}
The fixed charge arrangements of DCM can be further optimized by
releasing constraints on charge positions to exploit spatial degrees
of freedom, creating minimal distributed charge models.\cite{mdcm} For
this, MDCM employs machine-learning to determine charge positions and
magnitudes such that a predefined accuracy in the MEP is attained
using a minimal number of off-centered charges. Differential
Evolution\cite{storn1997DE} was found to be effective in this
context.\cite{mdcm} Unlike DCMs, MDCMs are not constrained to maintain
the same atomic multipole moments as a multipolar reference model, but
are fitted directly to describe the MEP around a molecule. Increasing
numbers of charges can be added until an MDCM representation reaches a
required level of accuracy, with the possibility to even improve
beyond what is possible with a multipolar model truncated at
quadrupole.\cite{mdcm} After fitting, each MDCM charge is assigned to
a nucleus and the MDCM arrangements are implemented in MD simulations
using the same framework (local axis systems, electrostatic cut-offs
etc.)  as a DCM model.\\

\subsection{Polarization}
For improved accuracy and physical rigour and realism, polarization
interactions were also included. This makes (M)DCM models viable for
emulating the iAMOEBA water model for which polarizable sites were
assigned to the atomic nuclei. Here, the `direct'
(non-self-consistent) approach was employed\cite{mccammon1990} which
allows direct comparison with the original iAMOEBA
model.\cite{wang2013systematic} In this approach, isotropic
polarizabilities at nuclear sites are used to add induced dipoles to
atoms as a function of the electric field generated by static
multipole moments of surrounding atoms only (the field generated by
other induced dipoles is ignored). For (M)DCM, this means that the
electric field is generated by the point charges of surrounding atoms
only. The total polarization energy is therefore:
\begin{eqnarray}
V_{\rm pol}=\sum_{i=1}^{N} \alpha_{i} {\rm \bf E}({\rm \bf
  r}_{i})^{2} \label{eq:epol1} \\ {\rm \bf E}({\rm \bf
  r}_{i})=\sum_{j=1}^{Nb_{i}}\sum_{l=1}^{Nq_{j}} \frac{\lambda_{ij}
  q_{l,j} {\hat{\rm \bf r}_{il}}}{R_{il}^2} \label{eq:epol2}
\end{eqnarray}
where the polarization energy $V_{\rm pol}$ is determined by a sum
over all $N$ atoms of their scalar (isotropic) polarizabilities
$\alpha_{i}$ multiplied by the square of the electric field ${\rm \bf
  E}$ at the atom's nuclear position ${\rm \bf r}_i$. The electric
field at the nuclear coordinate of atom $i$ is evaluated by summing
over each DCM charge $q_{l}$ of the $Nq$ DCM charges belonging to atom
$j$, for each atom in the list of $Nb$ nonbonded partners of atom $i$
within simulation cut-offs. ${\rm \bf \hat{r}}_{il}$ is a unit vector
in the direction of charge $l$ from polarizable center $i$, $R_{il}$
is the distance from the nucleus of atom $i$ to charge $l$. The
damping function $\lambda_{ij}$ used in
AMOEBA\cite{ren2003polarizable} is also adopted here, with functional
form:
\begin{eqnarray}
\lambda_{ij}=1-\exp \left( {-a \left(
  \frac{R_{il}}{(\alpha_{i}\alpha_{j})^{1/6}} \right)
  ^{3}}\right) \label{eq:epol3}
\end{eqnarray}
where $a = 0.23616$ \AA$^{-1}$.\cite{wang2013systematic} Although the
purpose of the damping function in the original AMOEBA force field was
to prevent artifacts at close range (the so-called `polarization
catastrophe'), in the non-iterative case these artifacts should not
exist, so $\lambda$ should be interpreted as a fitted short-range
correction to the polarization energy.

\subsection{Water Models}
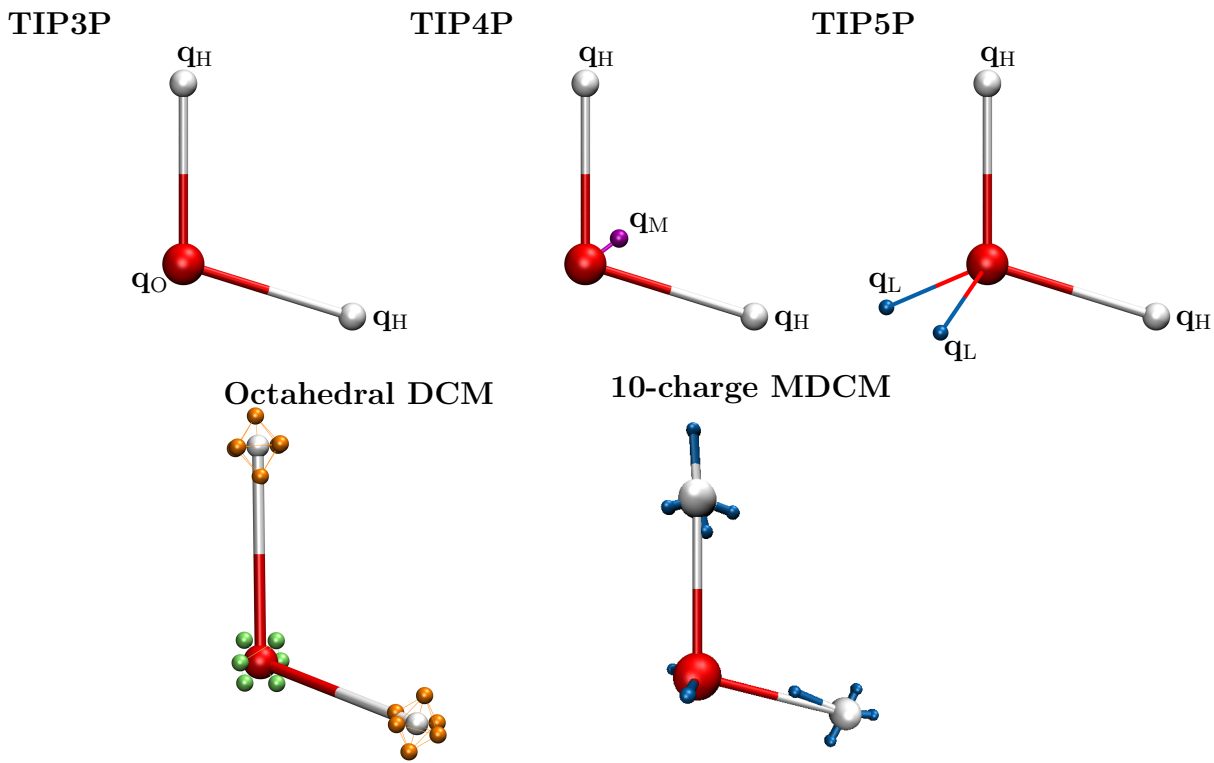
\begin{figure}[!ht]
\begin{minipage}[c]{5.2cm}
 \myhcolor{TIP3P}
\begin{tikzpicture}
    \node[anchor=south west,inner sep=0] (image) at (0,0) {\includegraphics[width=5.0cm,angle=0.0]{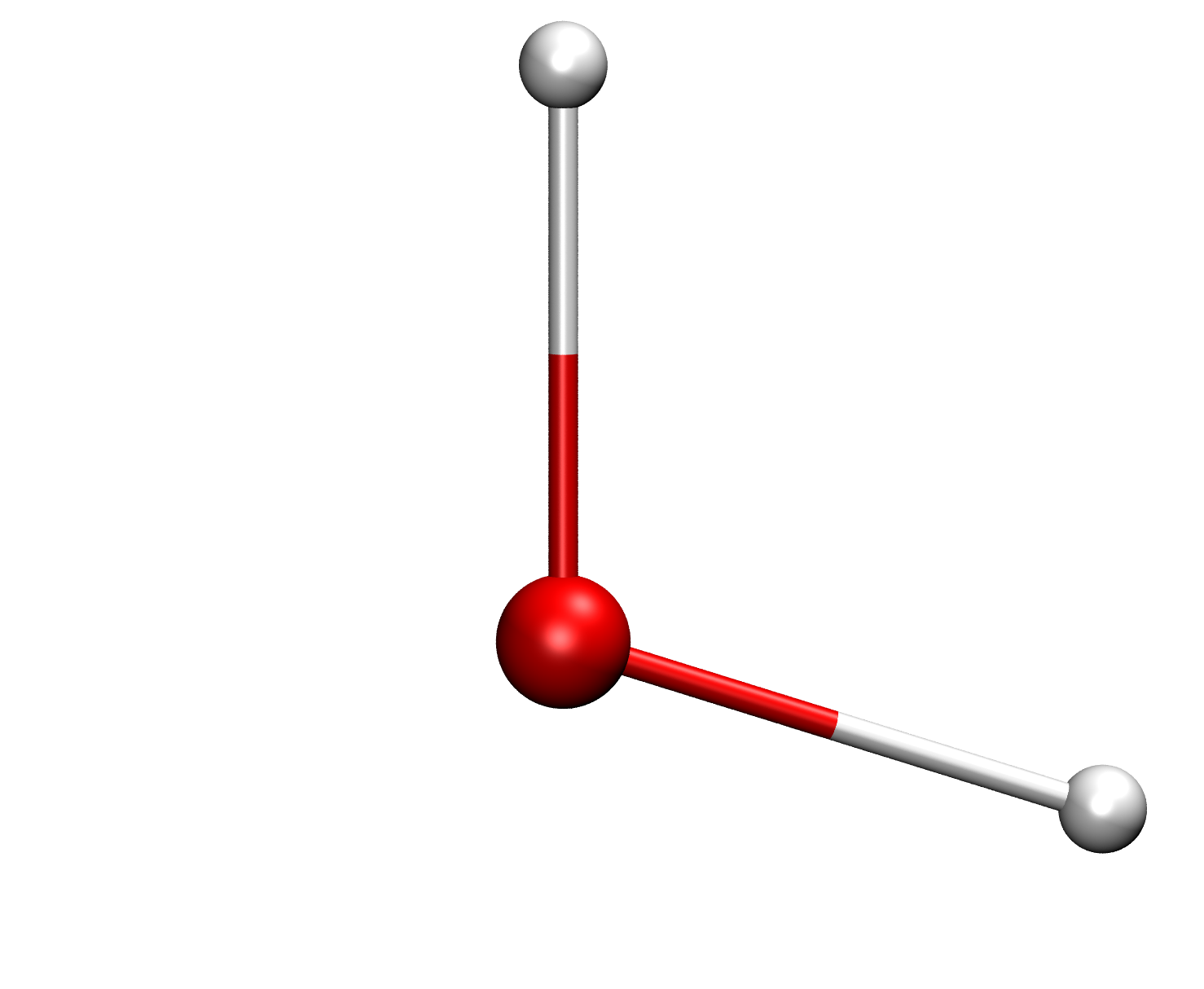}};
    \begin{scope}[x={(image.south east)},y={(image.north west)}]
    \end{scope}
    \node at (2.5,4.2) {{\bf\color{black} q$_{\rm H}$}};
    \node at (5.1,0.7) {{\bf\color{black} q$_{\rm H}$}};
    \node at (1.9,1.2) {{\bf\color{black}   q$_{\rm O}$}};
  \end{tikzpicture}
\end{minipage}
\begin{minipage}[c]{5.2cm}
 \myhcolor{TIP4P}
\begin{tikzpicture}
    \node[anchor=south west,inner sep=0] (image) at (0,0) {\includegraphics[width=5.0cm,angle=0.0]{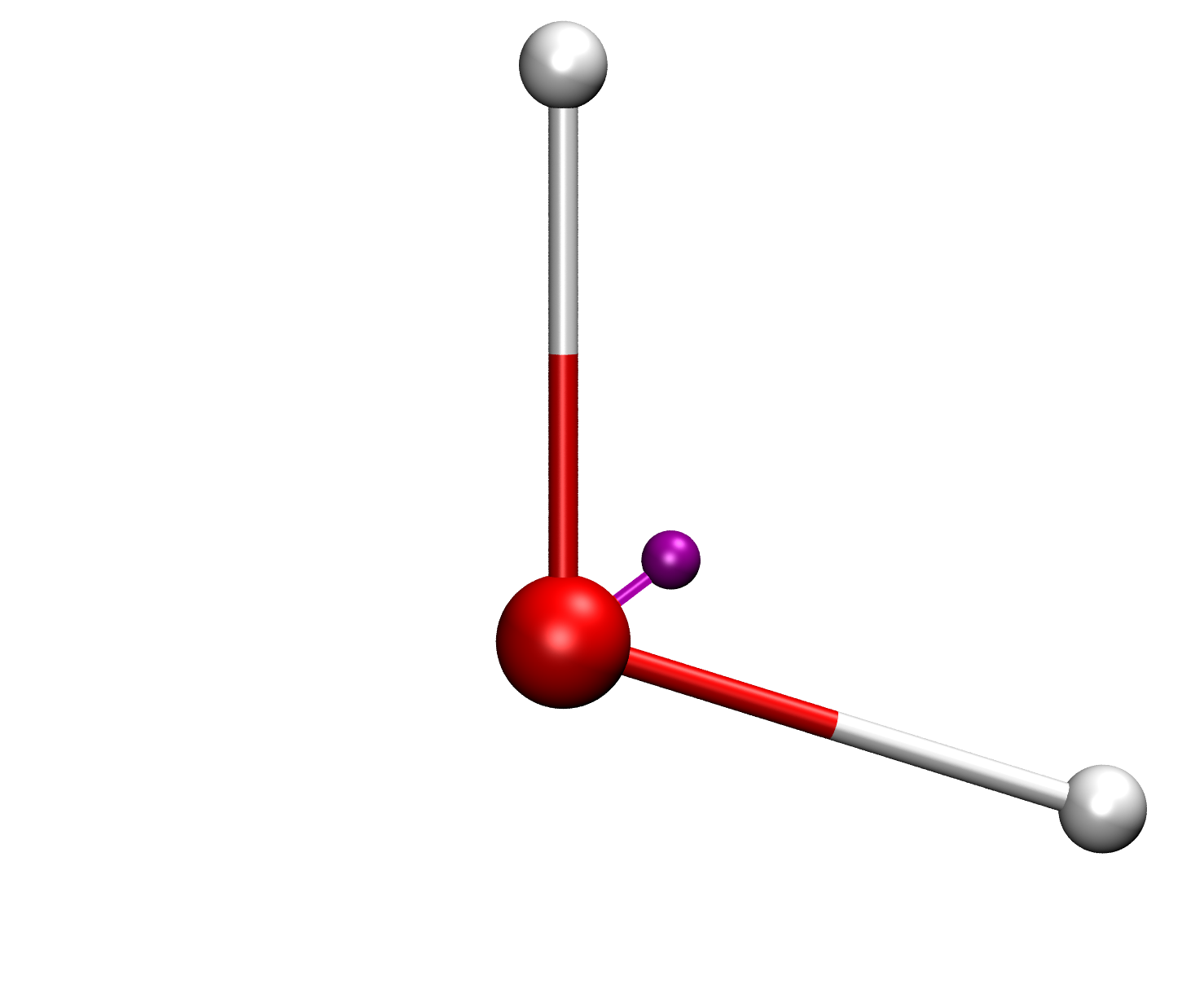}};
    \begin{scope}[x={(image.south east)},y={(image.north west)}]
    \end{scope}
    \node at (2.5,4.2) {{\bf\color{black} q$_{\rm H}$}};
    \node at (5.1,0.7) {{\bf\color{black} q$_{\rm H}$}};
    \node at (3.2,2.0) {{\bf\color{black}   q$_{\rm M}$}};
  \end{tikzpicture}
\end{minipage}
\begin{minipage}[c]{5.2cm}
 \myhcolor{TIP5P}
\begin{tikzpicture}
    \node[anchor=south west,inner sep=0] (image) at (0,0) {\includegraphics[width=5.0cm,angle=0.0]{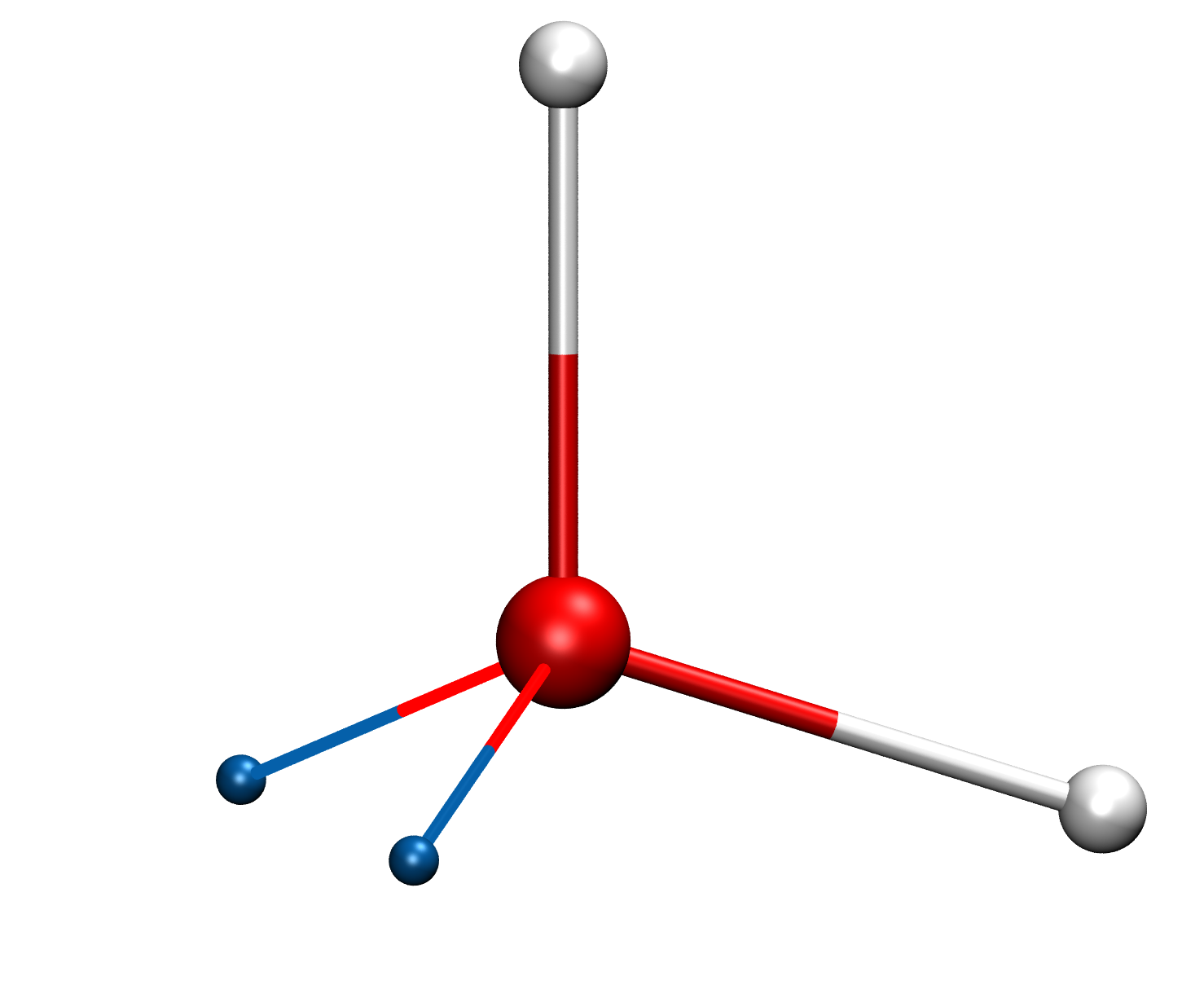}};
    \begin{scope}[x={(image.south east)},y={(image.north west)}]
    \end{scope}
    \node at (2.5,4.2) {{\bf\color{black} q$_{\rm H}$}};
    \node at (5.1,0.7) {{\bf\color{black} q$_{\rm H}$}};
    \node at (1.0,1.2) {{\bf\color{black}   q$_{\rm L}$}};
    \node at (2.0,0.3) {{\bf\color{black}   q$_{\rm L}$}};
  \end{tikzpicture}
\end{minipage} \\
\begin{minipage}[c]{5.0cm}
 \begin{flushleft}
 \myhcolor{Octahedral DCM}
  \begin{tikzpicture}
    \node[anchor=south west,inner sep=0] (image) at (0,0) {\includegraphics[width=3.0cm,angle=0.0]{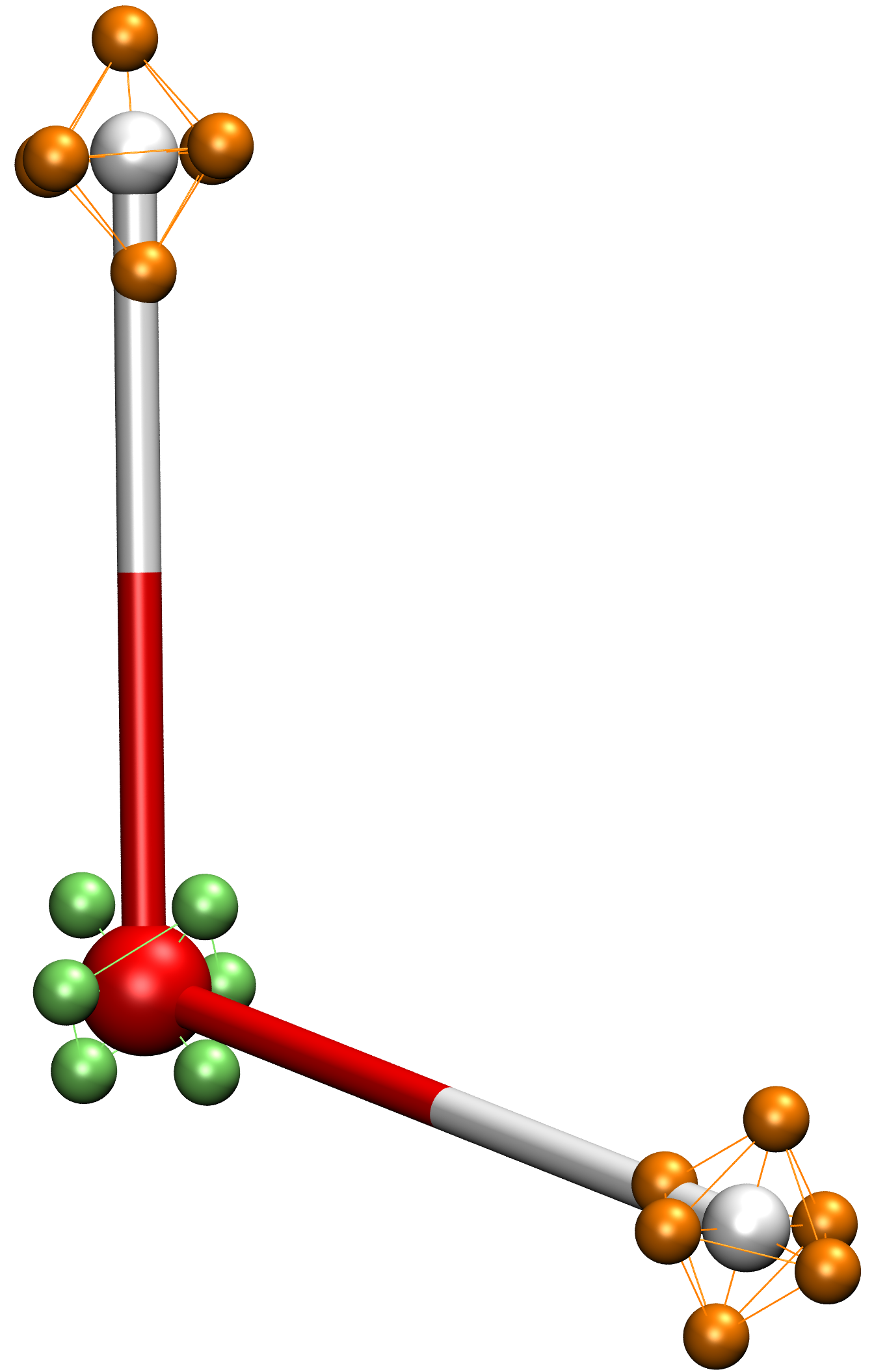}};
    \begin{scope}[x={(image.south east)},y={(image.north west)}]
    \end{scope}
  \end{tikzpicture}
  \end{flushleft}
\end{minipage}
\begin{minipage}[c]{5.0cm}
 \begin{flushleft}
 \myhcolor{10-charge MDCM}
  \begin{tikzpicture}
    \node[anchor=south west,inner sep=0] (image) at (0,0) {\includegraphics[width=4.0cm,angle=0.0]{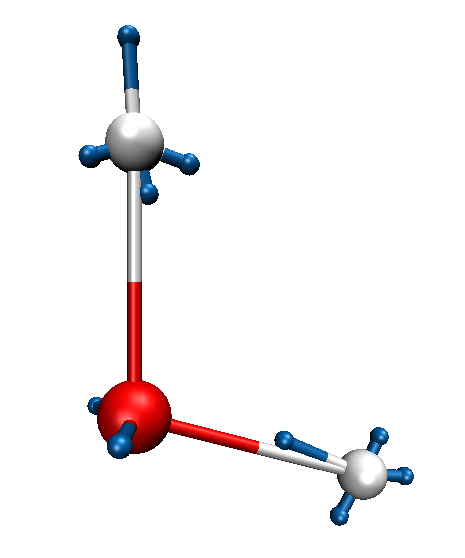}};
    \begin{scope}[x={(image.south east)},y={(image.north west)}]
    \end{scope}
  \end{tikzpicture}
  \end{flushleft}
\end{minipage}
\caption{Charge locations in TIP$n$P ($n=3,4,5$) water models (top),
  in an octahedral distributed charge model (DCM, bottom-left) and a
  10-charge minimal distributed charge model (MDCM, bottom-right). In
  all TIP$n$P models positive charges are located at the H
  centers. The corresponding negative charges are located either at
  the O atom ($q_{\rm O}$ in TIP3P), along an axis connecting O and
  the center of mass of H$_2$O ($q_{\rm M}$ in TIP4P), or at two
  points approximating the O lone pairs ($q_{\rm L}$ in TIP5P). In the
  octahedral DCM model there are 6 charge sites for the O atom and for
  each H atom, describing the multipole moments of the iAMOEBA
  model. In the MDCM model there are 4 charge sites per H atom and 2
  charge sites for O.}
\label{fig:tipnpgeom}
\end{figure}

{\bf TIP{\it n}P:} The parameters for these models are summarized in
Table \ref{tab:tipnpparam}. In all TIP{\it n}P models, the OH bond
length, $r_{\rm OH}$, and HOH bond angle, $\angle$HOH, are the
gas-phase experimental values, {\it i.e.}, 0.9572 \AA~ and
104.52$^\circ$, respectively. There is no charge at the O center in
both the TIP4P and TIP5P models. The potential energy of the TIP4P and
TIP5P models between two water molecules, $a$ and $b$, is given by
Eq. \ref{eq:ljpot}, where $i$ and $j$ are the charged sites on
molecules $a$ and $b$, respectively, and $r_{\rm O_{\rm a}O_{\rm b}}$
is the oxygen-oxygen distance.
\begin{align}
\label{eq:ljpot}
E_{\rm ab} &= 4\epsilon_{\rm OO}\left[ \left(\frac{\sigma_{\rm
      OO}}{r_{\rm O_{\rm a}O_{\rm b}}}\right)^{12} -
  \left(\frac{\sigma_{\rm OO}}{r_{\rm O_{\rm a}O_{\rm b}}}\right)^{6}
  \right] + \sum_{ij} \frac{q_i q_j}{r_{ij}}.
\end{align}
Eq. \ref{eq:ljpot} is equally applicable for TIP3P water when
including an additional L-J interaction term on hydrogen sites. As the
DCM approach uses off-centered charges to describe multipole moments,
no modification is necessary to implement the TIP$n$P models, which
are equivalent to MDCM distributions with 1 or 2 charges per atom.  In
some respects the TIP{\it n}P models can be viewed as MDCMs with
hand-fitted charge positions and magnitudes, as displayed in Figure
\ref{fig:tipnpgeom}.\\

\begin{table}[!ht]
\caption{
Parameters for TIP{\it n}P and iAMOEBA potential functions. 
}
\begin{center}
\begin{tabular}{lcrcrcrcrcr}
\hline
\hline
                                            && TIP3P    && TIP4P    && TIP5P  && \makecell{iAMOEBA\\(M)DCM}    \\
\hline                                                                                           
{\bf \color{black}Geometry}  \\                                                                  
$\angle$HOH$^{\circ}$                       && 104.52   && 104.52   && 104.52 && 106.48 \\
$\angle$LOL$^{\circ}$                       &&          &&          && 109.47 &&    \\
r$_{\rm OH}$/\AA                            && 0.9572   && 0.9572   && 0.9572 && 0.9584 \\
r$_{\rm OL}$/\AA                            &&          &&          && 0.70   &&   \\
r$_{\rm OM}$/\AA                            &&          && 0.15     &&        &&   \\
\hline                                                                                           
{\bf \color{black}Charges}  \\                                                                   
{\bf\color{black} q$_{\rm H}$}/{\it e}      &&  0.417   &&  0.52    &&  0.241 && $\dagger$   \\
{\bf\color{black}   q$_{\rm O}$}/{\it e}      && --0.834   &&          &&        && $\ddagger$  \\
{\bf\color{black}   q$_{\rm L}$}/{\it e}      &&          &&          && --0.241 &&  \\
{\bf\color{black}   q$_{\rm M}$}/{\it e}      &&          && --1.04    &&        &&  \\
\hline                                                                                           
{\bf \color{black}L-J parameters}  \\                                                            
$\epsilon_{\rm OO}$/kcal/mol         && --0.1521  && --0.155   && --0.16  && 0.19682$^a$  \\
$\epsilon_{\rm HH}$/kcal/mol         && --0.0460  &&          &&        &&  \\
$\sigma_{\rm OO}$/\AA                       &&  3.15061 &&  3.15365 &&  3.12  && 3.6453$^a$  \\
$\sigma_{\rm HH}$/\AA                       &&  0.4     &&          &&        &&  \\
\hline
{\bf \color{black} Polarizabilities}  \\
$\alpha_{\rm O}$/\AA$^{3}$ &&       &&          &&        &&   0.80636   \\
$\alpha_{\rm H}$/\AA$^{3}$ &&       &&          &&        &&   0.50484   \\
$a$/\AA$^{-1}$         &&       &&          &&        &&   0.23616$^{b}$   \\
\hline
\hline
\end{tabular}
\end{center}
\begin{flushleft}
$\dagger$ (M)DCM charge sites for H atoms, and $\ddagger$ (M)DCM
  charge sites for O atoms -- refer to section 2 of the SI.\\ $^{a}$
  iAMOEBA L-J parameters for use with a Halgren 14-7
  potential\cite{halgren1992representation} \\ $^{b}$ polarization
  damping factor
\end{flushleft}
\label{tab:tipnpparam}
\end{table}

{\bf iAMOEBA:} Inexpensive AMOEBA\cite{wang2013systematic} was
originally conceived as a computationally efficient (fewer parameters
and non-iterative polarization) and robust alternative to the existing
AMOEBA water model. The requisite Halgren buffered 14-7
potential\cite{halgren1992representation} and anharmonic bonded terms
of the iAMOEBA model are available in OpenMM,\cite{openMM} which also
contains basic ``dummy atom'' functionality for sites with zero mass
that can be used to run simulations with (M)DCMs. iAMOEBA in OpenMM
thus provides a suitable test case to compare performance of explicit
atomic multipole moments versus distributed charges in polarizable,
condensed phase simulations. It is also possible to demonstrate the
ease with which (M)DCMs can be made available in existing MD software
packages without the need to change the source code. Such an
``emulation'' should offer advantages in both computational efficiency
and ease of implementation with respect to adding explicit multipolar interaction terms to software that does not already have this functionality.\\

\section{Computational Details}
\subsection{DCM Representations}
TIP$n$P models were implemented for use with the DCM module in CHARMM
by describing the positions of any off-centered charges in the
standard DCM local axis system.\cite{devereux2014novel} For multipolar
models, in-house scripts were used to convert from the original local
axis system, as defined here by the AMOEBA force field for the iAMOEBA
model and the multipole module of the CHARMM force field for the
multipolar PhF model, to the final DCM axis system. Also, the
necessary diagonalization of the Cartesian quadrupole matrix to obtain
a minimal number of non-zero quadrupole components and to calculate
the corresponding charge magnitudes of the DCM arrangement were
computed using additional scripts that have been made available
online.\cite{MDCM-Git} An octahedral charge arrangement was used to
describe the multipole expansion for all DCM models, as described
previously.\cite{devereux2014novel} For the polarization term the
standard iAMOEBA damping parameters were used (Table
\ref{tab:tipnpparam}). Polarizabilities were corrected as described in
section \ref{Section:MDCMinOpenMM} below, to maintain the original
AMOEBA multipolar force field term and avoid refitting polarization
damping parameters. A sample input file is provided in section 3 of
the SI.\\

\subsection{MDCM fitting}
The Differential Evolution (DE) fitting code was implemented into the
``Fitting Wizard" (FW) tool previously developed to fit multipole
moments and L-J parameters to bulk properties for multipolar force
fields.\cite{fittingwizard2016} Charge positions were constrained
using hard constraints during DE fitting (candidate
solutions that violated constraints were labeled ``unfeasible'') so
that all charges remained within one third of the van der Waals radius
of an atom. Atomic multipoles up to rank $l=5$ were
fitted to the MEP across a grid using a least squares fit as
before,\cite{mdcm} where the grid used here is generated by the target
multipolar model of interest rather than quantum chemical reference
data. Specifically, the iAMOEBA and PhF multipole moments
were used to evaluate the MEP at each grid point of a rectangular grid
with spacing 0.1\ \AA\ and dimensions $11.4 \times 11.0 \times
10.1\ \text{\AA}$ (H$_{2}$O) or $14.4 \times 10.2 \times
15.2\ \text{\AA}$ (PhF) centered on the molecule.\\

\noindent
Grid points between the 0.001 a.u. and 0.0003 a.u. isodensity surfaces
were used for fitting, as points outside the outer 0.0003 a.u. surface
with lower electron density were found to be far enough away to be
generally well described and have only a small impact on fitting
quality. Points outside the outer surface that were excluded for
fitting were, however, included to validate the performance of the
model for the long-range part of the interaction. Grid points within
the 0.001 a.u. isodensity surface were discarded for both fitting and
subsequent evaluation of the fit.\\

\noindent
Atomic charge models with up to 4 charges per atom were fitted to the
ESP generated by the $l=5$ atomic multipoles for both molecules and used to
generate initial populations for subsequent DE fitting of the larger
systems, as described elsewhere.\cite{mdcm} MDCM was algorithmically
improved by introducing an intermediate fragmentation strategy to
increase computational efficiency for larger molecules. After fitting
atomic charge models to the ESP of the atomic multipoles, the PhF
molecule was thus divided into fragments (here into 2 arbitrary
fragments of roughly equal size). Next, each fragment was fitted
separately to a reference fragment ESP that does not include the ESP
contribution of the multipoles of the other fragment(s). This is
achieved using the high-rank ($l=5$) fitted atomic multipoles already
obtained to fit the atomic charge models:
\begin{eqnarray}
V_{\rm ref}^{\rm frag}({\bf r}) = V_{\rm ref}^{\rm mol}({\bf r}) - \sum_{i=1}^{N_{\rm frag,fix}} \sum_{j=1}^{N_{{\rm atom},i}} V_{i,j}^{\rm mtp}({\bf r})
\end{eqnarray}
where the fragment reference ESP $V_{\rm ref}^{\rm frag}$ at point
${\bf r}$ is equal to the original reference MEP, $V_{\rm ref}^{\rm
  mol}({\bf r})$, minus the ESP $V_{i,j}^{\rm mtp}({\bf r})$ from the
fitted multipoles of all $j$ atoms of all $N_{\rm frag,fix}$ fragments
that are not included in the current fragment fit. As each fragment
contains fewer charges than the full molecule, and all fragments can
be fitted independently a considerable speedup of the fitting process
is possible and the approach scales favourably for larger
systems. For H$_2$O no fragmentation was necessary and
molecular MDCMs were fitted directly to the iAMOEBA reference
MEP. Scripts and code required for this workflow have been made
available online.\cite{MDCM-Git}\\

\noindent
Fragments (PhF) or molecules (H$_{2}$O) are fitted with increasing
numbers of charges until a predefined/desired accuracy has been
obtained. Here, an average of between 1 and 3 charges per atom (PhF)
or 2 and 3.3 charges (H$_{2}$O), respectively, was trialled during
fitting and charges were free to move. The initial fragment or
molecular DE population is assembled from combinations of atomic
charge models that yield the lowest RMSE, typically by assigning more
charges to atoms with more challenging ESP distributions. Hence, the
number of charges for each atom within the fragment or molecule may
differ in the initial population, and may also change during fitting,
while the total number of charges for the fragment or molecule remains
fixed. For each given total number of charges ten independent models
for each PhF fragment or H$_{2}$O molecular model were fitted. After
fitting fragment models for PhF, those with lowest RMSE across the ESP
grid were combined to build molecular charge models with the total
number of charges also corresponding to between 1 and 3 charges per
atom (i.e. between 12 and 36 charges for the full PhF molecule) on
average. Each molecular PhF model was subjected to a final DE
refinement step.\\

\noindent
A further improvement to the original approach\cite{mdcm} was to
introduce additional constraints on charge magnitudes. Constraints are
important both to maintain stability of MD simulations and to maintain
accuracy of subsequent electrostatic interaction energy
calculations. Simulation stability is maintained by constraining
charge positions to remain within $r_{\rm atom} / 3$, one third of the
van der Waals radius of the atom. If charges are placed too far from
nuclear positions they are able to approach one another during MD
simulations, overcoming repulsive barriers and causing simulations to
collapse due to numerical instability. To improve the accuracy of
electrostatic interaction energies, hard constraints of maximally 1
$e$ for each point charge were applied to all charge magnitudes. The
grounds for constraining charge magnitudes is based on analysis of the
error in the interaction energy using MDCMs (described in section 4 of
the SI), as larger charge magnitudes often reduce the error in the MEP
at the expense of increasing the error in the electrostatic
interaction energy in subsequent simulations through error
multiplication. \\

\subsection{(M)DCM in OpenMM}
\label{Section:MDCMinOpenMM}
For iAMOEBA simulations, a single simulation engine (OpenMM
7.1.0\cite{openMM7}) was used for multipolar and (M)DCM models to keep all simulation details and force
field terms unchanged apart from the modified electrostatics. As
OpenMM lacks native DCM support, the existing ``dummy atom''
functionality was exploited to run (M)DCM simulations, highlighting
the possibility to run (M)DCM simulations in simulation packages that
support dummy atoms or equivalent features. (M)DCM charges were placed
relative to atoms by converting from local DCM axes to those defined
in OpenMM for dummy atoms (Figure \ref{fig:lra}). \\

\noindent
For consistency, the polarization term in OpenMM had to be adapted for
use with distributed charges. The polarization energy damping term
implemented in OpenMM assumes that charged and polarizable sites will
coincide, as the distance $R_{il}$ of the damping term in
Eq. \ref{eq:epol2} is evaluated between nuclear sites. This is not the
case in (M)DCM, where charge sites are shifted from nuclear
positions. If every (M)DCM charge site were assigned the
polarizability of the parent atom, the total polarization energy would
be significantly overestimated, in accordance with
Eq. \ref{eq:epol1}. Hence, $\alpha_{i}' = \alpha_{i} \cdot 10^{-4}$
was used for charge positions and the damping factor $\lambda_{ij}$
was changed to
\begin{eqnarray}
\lambda_{ij} = 1 - \exp{\left(-\frac{a'}{100} \left({\frac{r_{il}}{(\alpha_{i}'\alpha_{j}')^{1/6}}}\right)^{3}\right)}
\end{eqnarray}
with $a' = 0.0023616$ \AA$^{-1}$ and $\alpha_{i}' = \alpha_{i}$ for
the nuclei. As all nuclei carry zero charge in (M)DCMs the
nuclear--nuclear interactions yield zero polarization energy, in
accordance with Eq. \ref{eq:epol2}. As all charges carry
polarizabilities scaled by a factor $10^{-4}$, polarization energies
between (M)DCM charges are negligible which is consistent with
Eq. \ref{eq:epol1}. For charge--nuclear site interactions, the nuclear
site carries the standard polarizability, so Eq. \ref{eq:epol1} is
unchanged, and the charge site polarizability is scaled by $10^{-4}$,
which is counteracted by the factor $10^{-2}$ applied to the damping
parameter $a$ in Eq. \ref{eq:epol3}, recovering the polarization and
damping of the multipolar force field without refitting any
parameters. Note that the small remaining difference in the $R_{il}$
term from using shifted charge sites was found to not significantly
affect the results, but should be considered a potential source from
which slight differences can arise.\\

\noindent
Finally, fitted MDCMs were converted to the standard local axis
systems used for dummy atoms in OpenMM. A sample parameter file is
provided in section 5 of the SI.\\

\subsection{MD Simulations and Property Computation}
\label{Section:MDSimAndPropComp}
Aside from iAMOEBA simulations, which were run with OpenMM as
described below, remaining MD simulations were run with
CHARMM\cite{charmm} version 45a2 which includes provisions for
DCM.\cite{devereux2014novel} A 1 fs time step was used with
SHAKE\cite{shake-gunsteren} to constrain angles and bonds involving
hydrogen atoms in an isothermal-isobaric (constant $NPT$) ensemble
using a pressure bath at 1 atm coupled to a Nos\'e-Hoover temperature
bath\cite{andersen1980molecular,melchionna1993hoover,
  martyna1994constant,nose1984molecular,hoover1985canonical}. The
simulation system was a cubic box with 500 water molecules employing
periodic boundary conditions. For every value of $T$, a simulation of
at least 3 nanoseconds (ns) was performed. For TIP5P water and
temperatures below freezing point (273 K) the simulations were
extended by an additional 3 ns for improved estimates of thermodynamic
properties using fluctuation formulae (see below). This strategy has
been suggested previously to obtain converged results for modeling
bulk water at low
temperatures\cite{jorgensen1998temperature,mahoney2000five,baez1994existence,jorgensen2005potential}.
All simulations were performed with SHIFT and SWITCH cutoff
functions\cite{brooks-cutoffs} for non-bonded electrostatics and van
der Waals interactions, respectively. The switching-function
parameters are $R_{\rm on}$ and $R_{\rm off}$ with values 10.0 and
12.0 \AA, respectively, for non-bonded van der Waals interactions. A
12.0 \AA \ cutoff was applied for the shifted non-bonded
electrostatics. For atoms with off-centered charges,
distances for the shifting function were measured between the
off-centered charge sites. The TIP4P and TIP5P simulations in
particular provide a stringent validation for the implementation of
cut-offs, torques, and integration with barostats of (M)DCM into
CHARMM by comparing with results of the same simulations from those
using standard routines.\cite{Lonepair} PME for DCM has not yet been
implemented in CHARMM.\\

\noindent
OpenMM simulations of a cubic box with 500 water molecules were run
without SHAKE constraints for compatibility with iAMOEBA, and with a
0.5 fs time step. A Monte Carlo barostat maintained simulation
pressure at 1 atm, after 150 ps equilibration simulations were run for
10 ns at each $T$ used for the CHARMM simulations to facilitate direct
comparison between CHARMM and OpenMM data. Particle Mesh Ewald (PME)
was used with a real-space cutoff of 7.0 \AA \ and a van der Waals
cutoff of 9.0 \AA.\\

\noindent
{\bf Solvation Free Energies}, $\Delta G$, for PhF were calculated
using a thermodynamic integration (TI) procedure described
elsewhere,\cite{bereau2013} with slow-growth thermodynamic
integration\cite{kirkwood1935TI,Kollman1987SlowGrowth} for LJ
interactions with discrete windows of $\lambda$ from 0 to 1,
$\Delta\lambda = 0.1$ and $\Delta t = 1$ fs, averaging over 8 separate
forwards and backwards trajectories. $\lambda$ windows with an energy
variance of $> 0.5$ kcal/mol were halved and simulations were repeated
for the smaller window size. Although the DCM module in CHARMM is
integrated with existing slow-growth algorithms, for direct comparison
with multipole results the electrostatic contribution to the solvation
free energy was evaluated by performing simulations with solute
electrostatics scaled by $\lambda_{\rm mid}$, where $\lambda_{\rm
  mid}$ is the $\lambda$ value corresponding to the midpoint of the
$\lambda$ window, again for discrete windows of $\lambda$ from 0 to 1,
with fixed $\Delta\lambda = 0.05$ and $\Delta t = 1$ fs. A
post-processing step then recalculated energies using the unscaled
Hamiltonian for each time step of the simulation with scaled solute
electrostatics.\cite{bereau2013} The simulations for each value of
$\lambda$ for each of the 8 runs were equilibrated for 50 ps, and then
sampled for another 100 ps, {\it i.e.}, cumulatively $8 \times 150$ ps
for each TI window.\\

\noindent
{\bf Bulk-density $\rho$} was computed from the ratio between total
mass, $M$, and the time-averaged volume of the simulation box, $<V>$
according to $\rho = \frac{M}{<V>}$.\\

\noindent
{\bf Self-diffusion coefficient $D$} was computed from the mean
squared displacement (MSD) of all oxygen atoms using the Einstein
relation
\begin{align}
D &= \lim_{t \to \infty} \frac{1}{6t}<|r(t)-r(0)|^2>,
\end{align}
where $r(t)$ is the position of the oxygen atom of a water molecule at
time $t$, and averaged over all water
molecules\cite{mark2001structure}. OpenMM trajectories were analyzed
in CHARMM and $D$ was computed in the same way. \\

\noindent
{\bf Enthalpy of vaporization $\Delta H_{\rm vap}$} can be obtained
from
\begin{align}
\Delta H_{\rm vap} &= <E_{\rm gas}> - <E_{\rm liq}>/N +RT,
\end{align}
where $E_{\rm liq}$ is the potential energy of the liquid containing
$N$ molecules and $R$ is the ideal gas constant
\cite{jorgensen1998temperature,mahoney2000five,rick2004reoptimization}.\\

\noindent
{\bf Heat capacity $C_p$, isothermal compressibility $\kappa$,
  coefficient of thermal expansion $\alpha$} can be calculated from
standard fluctuation formulae (Eqs. \ref{eq:cp} to
\ref{eq:alpha})\cite{jorgensen1998temperature,mahoney2000five,rick2004reoptimization}.
\begin{align}\label{eq:cp}
C_{p}  &= \left(\frac{\partial H}{\partial T}\right)_{N, P}
        = \frac{1}{Nk_BT^2}(<H^2> - <H>^2) + 3R
\end{align}
\begin{align}\label{eq:kappa}
\kappa &= -\frac{1}{V}\left(\frac{\partial V}{\partial P}\right)_{N, T}
        = \frac{1}{k_BT<V>}(<V^2> - <V>^2)
\end{align}
\begin{align}\label{eq:alpha}
\alpha &= \frac{1}{V}\left(\frac{\partial V}{\partial T}\right)_{N, P}
        = \frac{1}{k_BT^2<V>}(<VH> - <V><H>)
\end{align}
Here, $C_p$ and $\alpha$ were computed using the central difference formula
for estimating derivatives, except at extremes where 
 right and left differences were used \cite{horn2004development},
\begin{align}\label{eq:cpalpha}
C_{p}  \approx \frac{<H_2> - <H_1>}{T_2-T_1} \quad , {\rm and} \quad
\alpha \approx \frac{{\rm ln}<\rho_2> - {\rm ln}<\rho_1>}{T_2-T_1}
\end{align}
and $\kappa$ values were calculated from the fluctuations
(Eq. \ref{eq:kappa}).\\

\section{Results and Discussion}
This section is structured as follows. First, the performance of the
(M)DCM iAMOEBA model is evaluated for single-point energy calculations
of water clusters, in order to validate the (M)DCM models against
reference data from the original multipolar iAMOEBA
implementation. Implementations for the TIP$n$P models are straightforward as the original charge coordinates are used directly, rather than replaced by an (M)DCM, and
are hence not shown here. Then the performance of all models for
different bulk properties between $T=235.5$ K and $T = 350$ K is
assessed to explore the performance of the DCM framework in describing
models of increasing complexity under realistic simulation
conditions. For iAMOEBA the impact on simulations of replacing the
multipole term with distributed charges is the main focus. For TIP$n$P the focus is compatibility of the DCM dynamics framework with existing, simpler charge models with and without off-centered charges. Finally,
the methods developed are further used for MD and free energy
simulations of hydrated PhF.\\

\subsection{MDCM Fitting for H$_{2}$O}
Models with increasing numbers of charges ranging from 6 to 10 per
molecule were fitted to the MEP, with 10 independent models fitted for
each number of charges. The models that performed best as quantified
by mean absolute error in the MEP were then used in dimer energy
calculations (next section). As no refitting of other force field
parameters was desirable a tight agreement of $\sim 0.2$ kcal/mol (1
kJ/mol) in dimer energies was the threshold to select the MDCM to be
used in subsequent calculations. It was found that 10 charges were
required to reach this threshold, more than might typically be
necessary for such a model. However, this is justified by the
challenging nature of the task, and still compares favorably with the
12 non-zero multipole components of the iAMOEBA model and without the
additional complexity that these terms will incur in simulations. In a
more typical application an error in dimer electrostatic interaction
energies of 1 kcal/mol with respect to {\it ab initio} reference data
might be acceptable, requiring fewer charges, as errors are typically
compensated by remaining force field terms. The MAE across the grid
between the two isodensity surfaces for the selected model was 0.023
kcal/mol and the maximum absolute error was 2.10 kcal/mol.\\

\noindent
The selected model is shown at the bottom-right of Figure
\ref{fig:tipnpgeom}. No symmetry constraints were applied in the
  fitting. While such constraints would represent a useful extension
for the particular case of systems with symmetry, the asymmetric
distribution of MDCMs obtained during fitting still accurately
describes the symmetry of the underlying MEP if the fitting criteria
are sufficiently tight. This is explicitly discussed for the
  water cluster energies in section S6 of the SI. There, it is
  demonstrated that minimal error is incurred by rotating monomers
180 degrees about their H--O--H bisectors, with maximal difference of
$\Delta E \leq 0.027$ kcal/mol for dimer energies and $\Delta E \leq
0.019$ kcal/mol/monomer for clusters up to decamer, and mean absolute
$\Delta E = 0.039$ kcal/mol over all dimers and clusters. Both
orientations can therefore be used equivalently in simulations without
affecting bulk properties, as further demonstrated in sections
\ref{section:298K} to \ref{section:PhFProps} below.

\begin{figure}[!ht]
  \includegraphics[width=8.0cm]{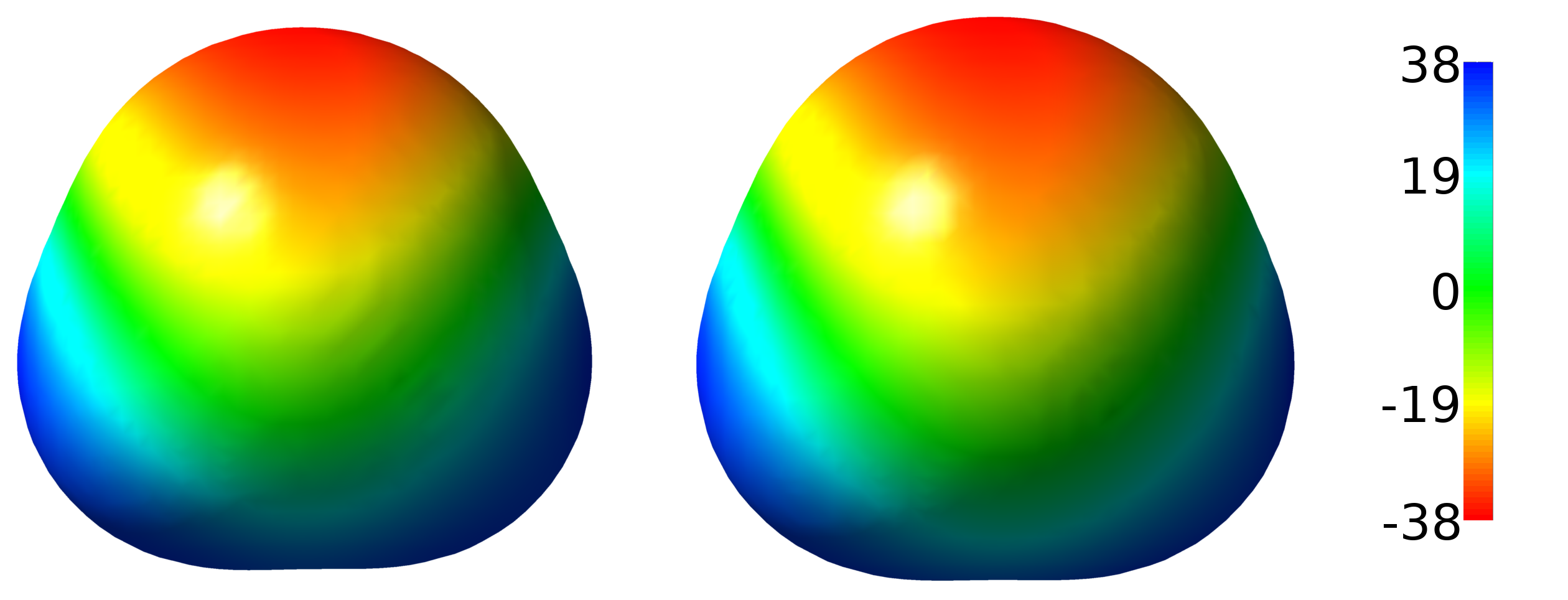}\\
  \includegraphics[width=10.0cm]{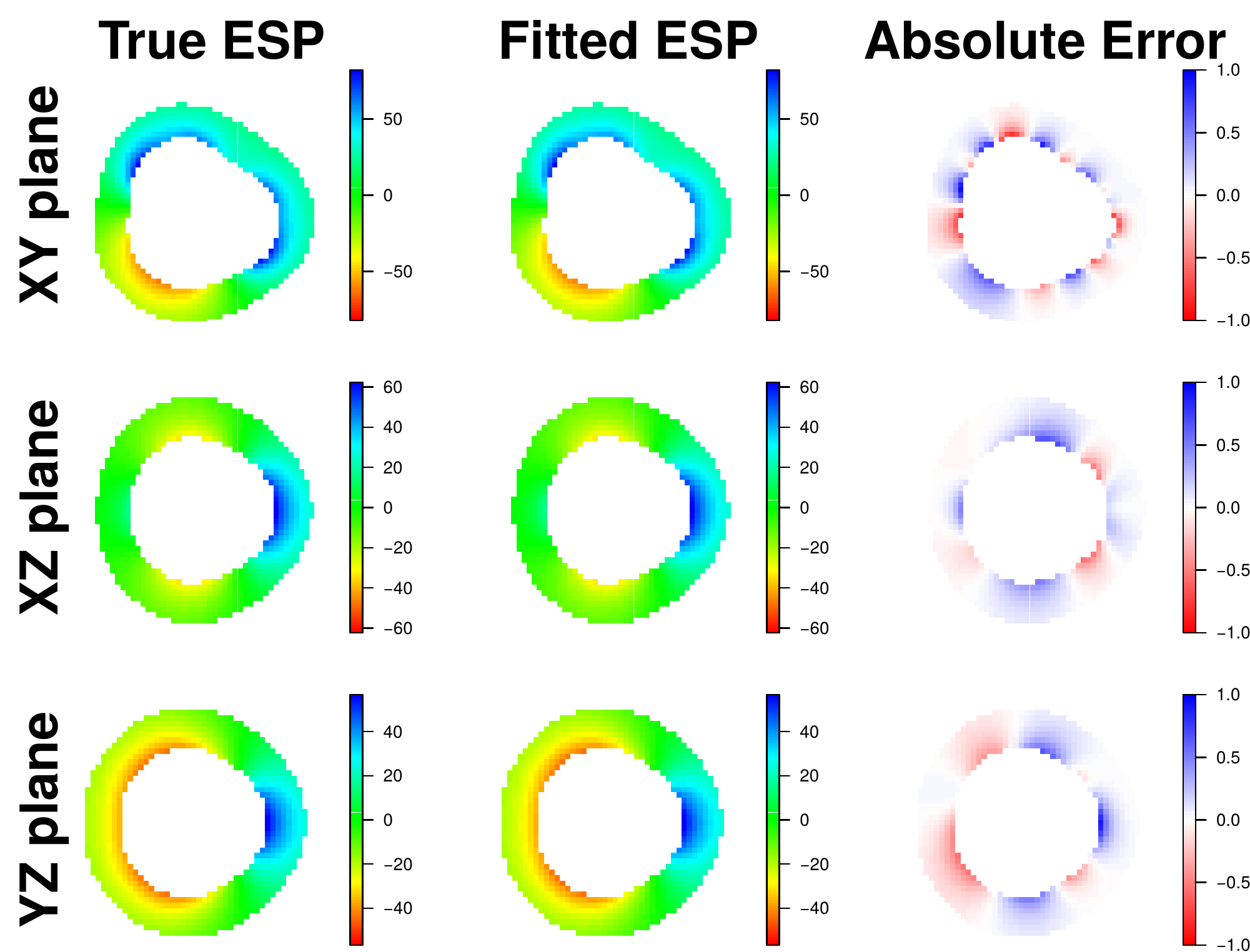}
\caption{Top: comparison of MEP generated by 10-charge MDCM (left)
  with MEP generated by iAMOEBA multipole moments (right), mapped onto
  the molecular 0.001 a.u. isodensity surface. Colors range from -38
  (red) to +38 (blue) kcal/mol. Bottom: 2D slices of the MEP in
  different molecular planes (rows) in the region used for
  fitting. The ``True'' ESP refers to the iAMOEBA multipolar
  reference, the fitted ESP refers to the MDCM 10-charge model. The
  absolute error is plotted in the right-hand column and ranges from
  -1 kcal/mol (red) to +1 kcal/mol (blue).}
\label{fig:h2o_esp_fit}
\end{figure}

\subsection{iAMOEBA and its (M)DCM Representation}
\begin{figure}[!ht]
\includegraphics[width=8.0cm]{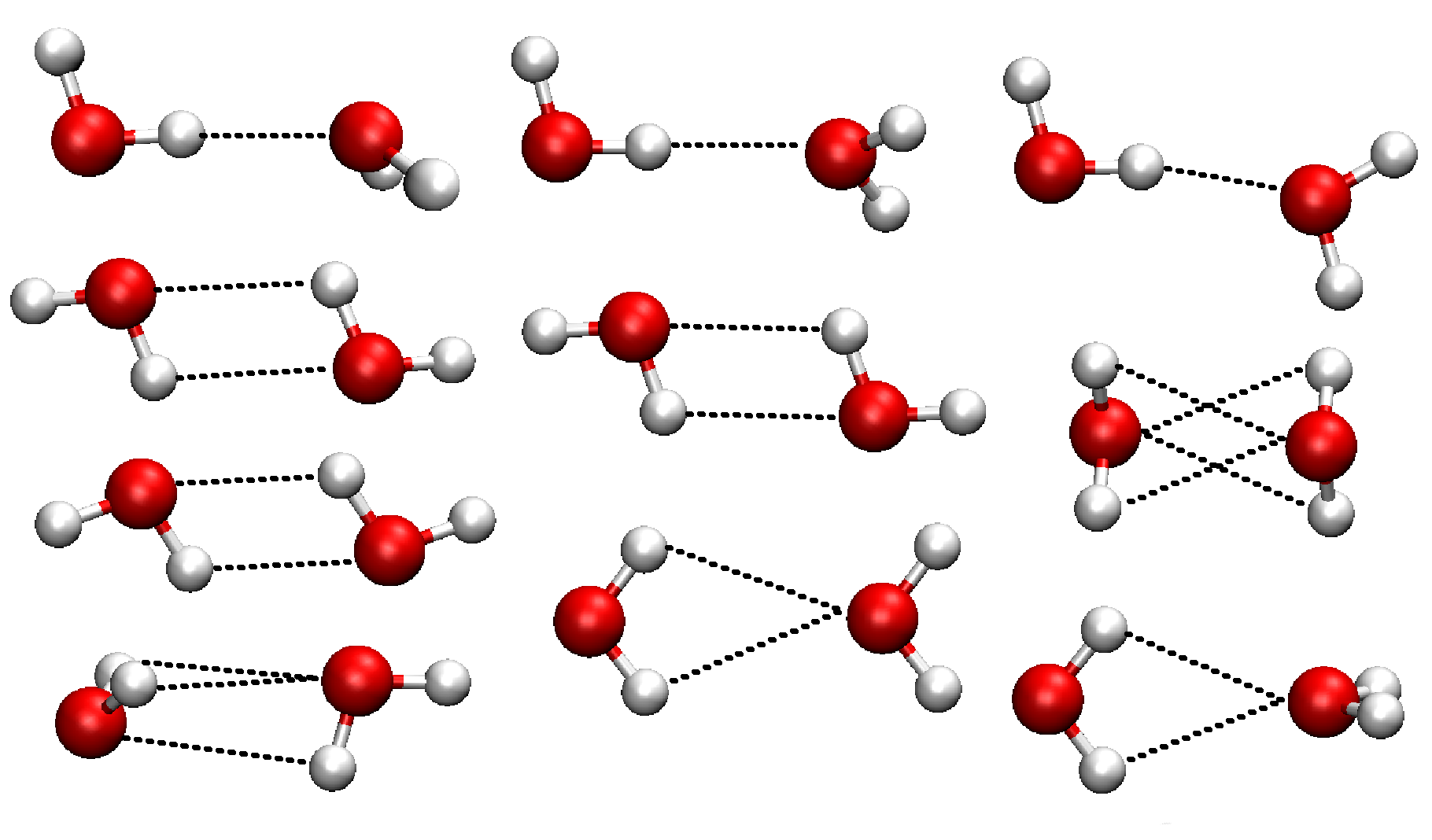}
\caption{Water dimers from Ref. \cite{tschumperDimers} used to
  validate (M)DCM iAMOEBA parameters.}
\label{fig:dimers}
\end{figure}

The quality of the octahedral DCM and 10-charge MDCM descriptions of
the iAMOEBA multipolar electrostatics were examined using a series of
water dimers, originally proposed by Tschumper {\it et
  al.}\cite{tschumperDimers} (Figure \ref{fig:dimers}) and a set of
larger water clusters up to decamer\cite{WalesClusters} to check for
error accumulation with cluster size.\\

\begin{figure}[!ht]
\includegraphics[width=9.0cm]{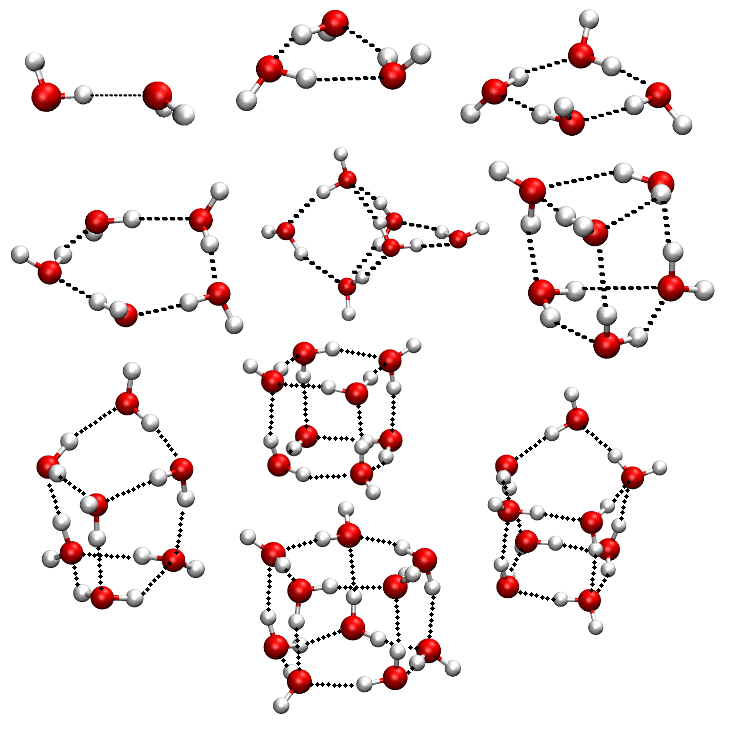}
\caption{ Water clusters used to check for error accumulation in (M)DCM
  iAMOEBA model. }
\label{fig:clusters}
\end{figure}

\begin{table}
\caption{
  Comparison between multipolar and DCM iAMOEBA electrostatic
  energies, including polarization. In the top part of the table,
  energies for the 10 dimer structures (see Figure \ref{fig:dimers})
  are shown, while in the bottom part energies for various oligomers
  (see Figure \ref{fig:clusters}) are presented. The mean absolute
  error for the dataset is within chemical accuracy (MAE=0.041
  kcal/mol)}
\begin{tabular}{l |c*{4}{c}r}
Dimer  &iAMOEBA (kcal/mol)&DCM-iAMOEBA (kcal/mol)& $\Delta E$ (kcal/mol)\\
\hline
1	&	--5.113	&	--5.092	&	  0.021	\\
2	&	--4.505	&	--4.532	&	--0.027	\\
3	&	--4.504	&	--4.554	&	--0.049	\\
4	&	--3.824	&	--3.769	&	  0.056	\\
5	&	--3.269	&	--3.217	&	  0.053	\\
6	&	--2.972	&	--2.967	&	  0.005	\\
7	&	--3.199	&	--3.173	&	  0.027	\\
8	&	--1.572	&	--1.579	&	--0.007	\\
9	&	--3.794	&	--3.778	&	  0.016	\\
10	&	--3.016	&	--3.027	&	--0.011	\\
\hline
Oligomer  &iAMOEBA (kcal/mol)&DCM-iAMOEBA (kcal/mol)& $\Delta E$ (kcal/mol)\\
\hline
trimer	&	--13.770	&	--13.691	&	0.079	\\
tetramer	&	--24.529	&	--24.495	&	0.034	\\
pentamer	&	--32.321	&	--32.279	&	0.067	\\
hexamer prism	&	--41.459	&	--41.421	&	0.038	\\
heptamer	&	--51.299	&	--51.193	&	0.106	\\
octamer	&	--64.672	&	--64.495	&	0.178	\\
nonamer	&	--72.896	&	--72.989	&	--0.092	\\
decamer	&	--82.858	&	--82.772	&	0.086	\\
\label{tab:dcm-clusters}
\end{tabular}
\end{table}

\begin{table}
\caption{Comparison between multipolar and MDCM iAMOEBA electrostatic
  energies, including polarization. In the top part of the table,
  energies for the 10 dimer structures are shown, while in the bottom
  part energies for various oligomers are presented. The mean absolute
  error for the dataset is within chemical accuracy (MAE=0.108
  kcal/mol)}
\begin{tabular}{l |c*{4}{c}r}
Dimer  &iAMOEBA (kcal/mol)&MDCM-iAMOEBA (kcal/mol)& $\Delta E$ (kcal/mol)\\
\hline
1 & --5.113 & --5.125 & --0.011\\
2 & --4.505 & --4.465 & 0.040\\
3 & --4.504 & --4.471 & 0.033\\
4 & --3.824 & --3.836 & --0.011\\
5 & --3.269 & --3.180 & 0.089\\
6 & --2.972 & --2.863 & 0.109\\
7 & --3.199 & --3.196 & 0.003\\
8 & --1.572 & --1.587 & --0.014\\
9 & --3.794 & --3.770 & 0.024\\
10 & --3.016 & --3.026 & --0.010\\
\hline
Oligomer  &iAMOEBA (kcal/mol)&MDCM-iAMOEBA (kcal/mol)& $\Delta E$ (kcal/mol)\\
\hline
trimer	&	--13.770	&	--13.782	&	--0.012	\\
tetramer	&	--24.529	&	--24.436	&	0.093	\\
pentamer	&	--32.321	&	--32.266	&	0.055	\\
hexamer prism	&	--41.459	&	--41.348	&	0.111	\\
heptamer	&	--51.299	&	--51.129	&	0.171	\\
octamer	&	--64.672	&	--64.238	&	0.434	\\
nonamer	&	--72.896	&	--72.478	&	0.419	\\
decamer	&	--82.858	&	--82.402	&	0.456	\\
\label{tab:mdcm-clusters}
\end{tabular}
\end{table}

\noindent
As seen in Table \ref{tab:dcm-clusters}, very close agreement was
obtained between the original multipolar iAMOEBA implementation in
OpenMM and its DCM representation developed here. The difference in
electrostatic interaction energy (including polarization) is of the
order of $10^{-2}$ kcal/mol for all dimers. Close agreement was also
obtained for the water clusters, with error accumulation remaining
remarkably small up to decamer, and the largest total error of 0.18
kcal/mol still well within chemical accuracy.\\

\noindent
Table \ref{tab:mdcm-clusters} reveals a similar trend for the fitted
MDCM. While dimer errors are slightly larger than for the DCM, the
largest error of 0.109 kcal/mol is still very close to the iAMOEBA
multipolar energy. For the larger clusters results are again
encouraging, although some error accumulation is visible as errors
increase to almost 0.5 kcal/mol for the decamer. It should be
emphasized that this remains well within chemical accuracy even for
these larger clusters, with a percentage error of 0.5\% for the
decamer and MAE of 0.11 kcal/mol for the whole set of dimers and
clusters, despite requiring little more than half the number of
charges used in the octahedral DCM.\\

\subsection{H$_{2}$O Thermodynamic Properties at 298 K from all Models}
\label{section:298K}
After establishing that both the octahedral DCM and 10-charge MDCM
yield accurate interaction energies, the performance of the models for
bulk properties in condensed phase MD simulations was assessed. (M)DCM
routines and axis systems for models ranging from TIP3P to iAMOEBA and
(M)DCM charge representations of iAMOEBA were chosen to demonstrate
the versatility of the approach. For multipolar models, simulations
are carried out for both the original model and its (M)DCM
representation. For TIP$n$P models, simulations are carried out using
both preexisting code with routines for off-centered charges, and the
new DCM implementation in CHARMM.\\

\noindent
Atom-atom pair correlation functions (radial distribution functions,
RDFs) $g_{\rm OO}(r)$, $g_{\rm OH}(r)$ and $g_{\rm HH}(r)$ characterize the microscopic structure of liquid water. One of the
critical tests for water models is accurate reproduction of the
experimental (X-ray scattering and neutron diffraction) OO, OH and HH
RDFs, although it should be noted that there remains some uncertainty
over reference values in the literature. For example, the reported
height ($g1$) and position ($r1$) of the first intermolecular peak in
$g_{\rm OO}(r)$, which is the characteristic feature of liquid water,
varies over the range $g1 = 2.2$--3.0 and $r1 = 2.76$--2.82 \AA\/
depending on the type of
experiment.\cite{soper2000radial,hura2000high,fu2009x,leetmaa2008diffraction,
  wikfeldt2010oxygen,petkov2012molecular,hura2003water,head2002water,
  soper2007joint,skinner2013benchmark}\\

\begin{figure}[!ht]
\centering
\begin{minipage}[b]{0.9\textwidth}
\includegraphics[height=0.8\textwidth,angle=-90.0]{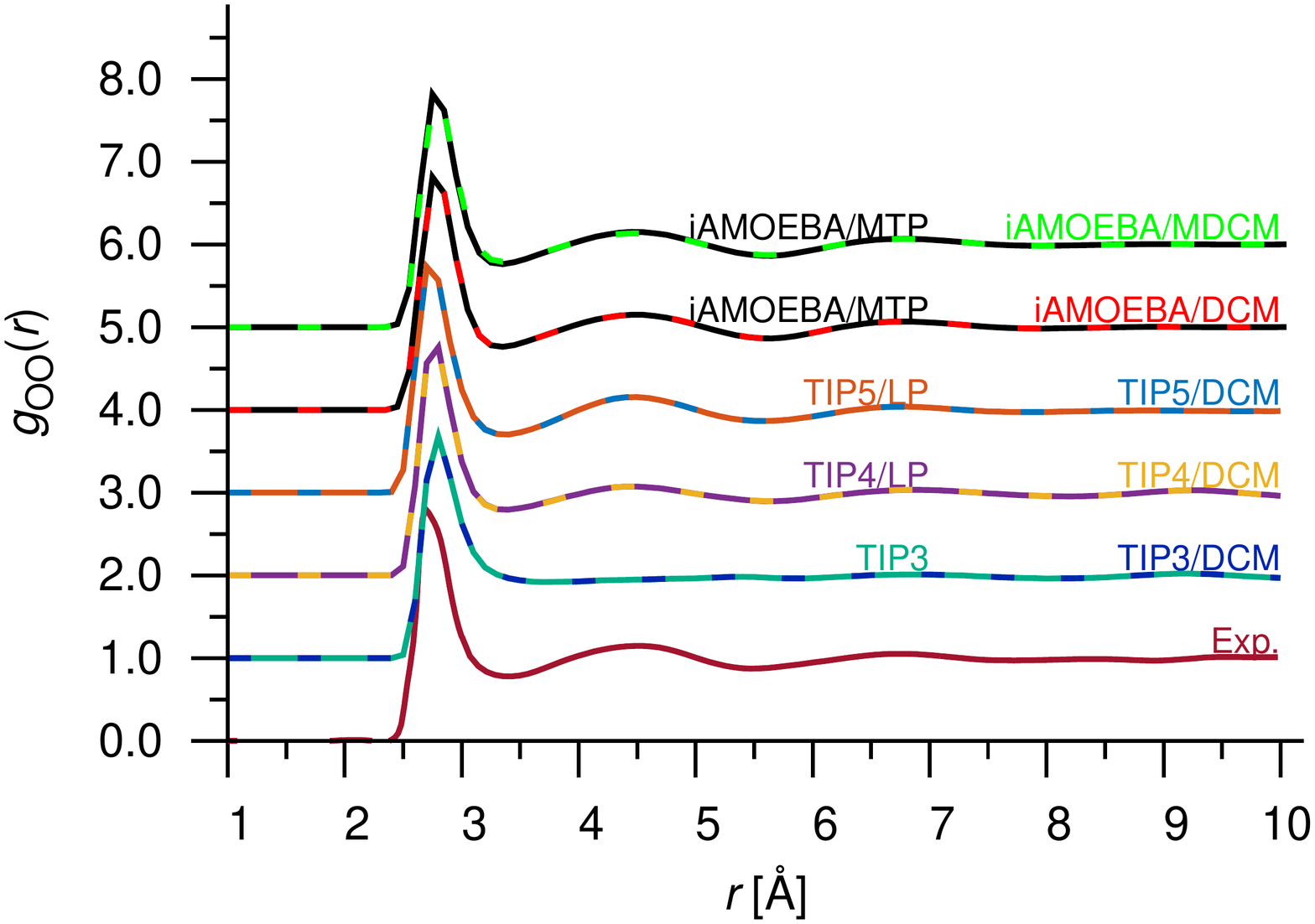}
\caption{Radial distribution functions (RDF) $g_{\rm OO}(r)$ of
  iAMOEBA and TIP{\it n}P H$_{2}$O at 298 K and 1 atm using standard
  routines (LP) and DCM functionality in CHARMM (TIP$n$P), and
  multipolar (MTP) iAMOEBA and (M)DCM implementations of iAMOEBA in
  OpenMM. Experimental neutron diffraction data from
  Ref. \cite{soper2000radial}. Successive curves are offset 1 unit
  along the $y$-axis for clarity.}
\label{fig:goor}
\end{minipage}
\end{figure}

\noindent
Figure \ref{fig:goor} shows $g_{\rm OO}(r)$ for the various water
models along with experimental neutron diffraction data
\cite{soper2000radial}. A more detailed comparison of $g_{\rm OO}(r)$
along with $g_{\rm OH}(r)$ and $g_{\rm HH}(r)$ is presented in
sections 7--9 of the SI. Of particular interest here is the close
agreement firstly between TIP$n$P results and their DCM
equivalents. The DCM implementation in CHARMM was conceived to
efficiently run dynamics for distributed charge representations of
multipolar electrostatic models. Decisions such as how to implement
cut-offs and torques, and interactions with barostats may therefore
yield different results to widely used dummy-atom or lone-pair
routines commonly used to implement TIP4P and TIP5P. The similar
behavior of both models using the DCM implementation and more widely
used approaches therefore both validates the code and demonstrates the
equivalence of the implementation in this context.
Secondly, there is very good agreement for the polarizable, multipolar
iAMOEBA results with (M)DCM, demonstrating that the close agreement in
MEP and interaction energies yields correspondingly close $g(r)$ in
simulations without the need for explicit multipoles, and using only
generic functionality for off-centered charge sites in OpenMM.\\

\noindent
Liquid density, heat of vaporization, isobaric heat capacity,
isothermal compressibility, thermal expansion coefficient and
self-diffusion coefficient at 298 K and 1 atm are summarized in Table
\ref{tab:bulk_prop}. Again results for TIP{\it n}P models are compared
with their DCM-representations and those for multipolar iAMOEBA are
compared with those from (M)DCM and with experiment.\\

\noindent
The calculated $\rho$, $\Delta H_{\rm vap}$ and $D$ of TIP{\it n}P
waters are essentially identical from both sets of simulations. $C_p$,
$\kappa$ and $\alpha$ values of water from the various models vary
within $\pm 1$ unit, although for TIP4P $\alpha$ is somewhat
overestimated, by about 30 \%. As shown in the next section, this
discrepancy seems to originate from noise in the data rather than a
physical effect, as performance across a range of temperatures shows
closer agreement and the other models also show local discrepancies at
certain temperatures (Figure \ref{fig:propsvst}c).  Agreement for the
simulations with iAMOEBA and its (M)DCM representations is also very
good, with the largest discrepancy being the slightly larger value for
$\kappa$ with the DCM representation. Figure \ref{fig:propsvst}b again shows this discrepancy is significantly smaller than the difference between different models.\\

\noindent
The range of dynamic properties examined provides confidence that the
microscopic structure and dynamics of the solvent are preserved when
moving from a full multipolar description to the simpler distributed
charge models, also after reducing the number of charges in the
MDCM. All iAMOEBA results additionally agree well with experiment,
consistent with earlier findings.\cite{wang2013systematic} It should
also be noted that the data presented are for the simulation
conditions described above, and hence certain deviations from
previously published results are to be expected. For example $\Delta
H_{\rm vap}$ is 6 \% larger than the originally published Monte Carlo
(MC) data\cite{jorgensen1983comparison}, $\kappa$ values are roughly
30 ($10^{6}$ atm$^{-1}$) smaller for TIP3P and TIP4P compared to
original MC
data\cite{jorgensen1983comparison,jorgensen1998temperature}, and $D$
for both TIP3P and TIP4P is roughly 40 \% larger than previously
published results.\cite{mahoney2001diffusion} As has been highlighted
previously,\cite{mark2001structure} small differences in simulation
conditions can have a significant impact on the results from bulk
simulations. \\

\begin{table}[!ht]
\caption{Bulk properties of liquid water at 298 K and 1 atm; density
  $\rho$ (g cm$^{-3}$), enthalpy of vaporization $\Delta H_{\rm vap}$
  (kcal/mol), isobaric heat capacity $C_p$ (cal mol$^{-1}$K$^{-1}$),
  isothermal compressibility $\kappa$ (10$^6$ atm$^{-1}$), thermal
  expansion coefficient $\alpha$ (10$^5$ K$^{-1}$) and self-diffusion
  coefficient $D$ (10$^{-5}$cm$^2$ s$^{-1}$). ``DCM" denotes the new
  DCM code and framework were used in place of standard routines.}
\begin{center}
\begin{tabular}{lcccccccc}
\hline
\hline
              & $\rho$ & $\Delta H_{\rm vap}$ & $C_p$  & $ \kappa$ & $\alpha$ & $D$    \\
\hline                                                          
TIP3P$^c$     & 1.0266 & 11.04             & 12.9   & 23.1      & 76.8     & 3.9  \\
TIP3P/DCM     & 1.0264 & 11.04             & 12.9   & 22.4      & 75.6     & 3.9  \\
\hline
TIP4P      & 1.0090 & 11.15             & 15.6   & 22.7      & 35.3     & 2.3  \\
TIP4P/DCM     & 1.0082 & 11.16             & 15.5   & 23.3      & 44.9     & 2.3  \\
\hline
TIP5P      & 0.9848 & 10.73             & 24.2   & 28.8      & 30.9     & 2.2  \\
TIP5P/DCM     & 0.9842 & 10.73             & 24.8   & 30.0      & 30.0     & 2.2  \\
\hline
iAMOEBA/OpenMM & 0.9977 & 10.91             & 17.8   & 40.7      & 23.3     & 1.9  \\
iAMOEBA/DCM   & 0.9916  &  10.85            & 18.3   & 51.0         & 25.8     & 2.0  \\
iAMOEBA/MDCM10  & 0.9964 &  10.76           & 17.9   & 41.5      & 28.9     & 2.2  \\
\bf Exp.$^a$      & \bf 0.9965 & \bf 10.51  & \bf 18.0   & \bf 45.8      & \bf 25.7     & \bf 2.3$^b$ \\
\hline
\hline
\end{tabular}\\
$^a$ Ref. \cite{wagner2002iapws}; 
$^b$ Ref. \cite{mills1973self};
$^c$ using conventional TIP3P in CHARMM;
\end{center}
\label{tab:bulk_prop}
\end{table}

\subsection{Temperature Dependence of H$_{2}$O Thermodynamic Properties}
\begin{figure}[!ht]
\centering
\begin{minipage}[b]{0.9\textwidth}
\includegraphics[width=0.8\textwidth,angle=-90]{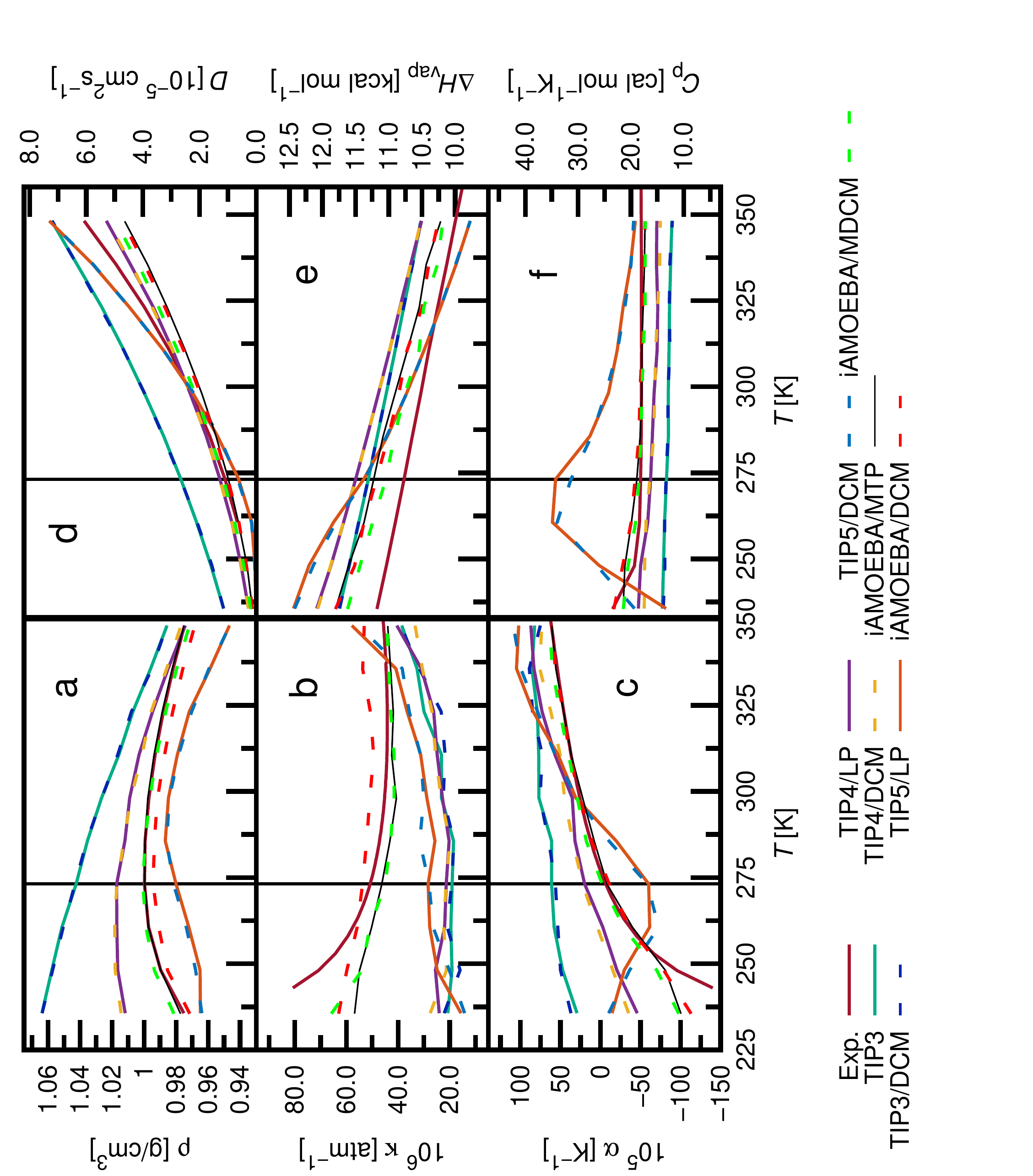}
\caption{Experimental and computed densities $\rho$, diffusion
  coefficients $D$, isothermal compressibilities $\kappa$, enthalpies
  of vaporization $\Delta H_{\rm vap}$, thermal expansion coefficients
  $\alpha$ and heat capacities $C_{\rm p}$ of liquid water using
  TIP{\it n}P and iAMOEBA water models as a function of temperature at
  1 atm Exp. data from
  Refs. \cite{archer2000thermodynamic,wagner2002iapws,kell1975density,mills1973self,gillen1972self}. Experimental
  $\rho$ data are obscured by the iAMOEBA/MTP (multipolar iAMOEBA)
  data. A vertical line at 273 K indicates zero Celsius.}
\label{fig:propsvst}
\end{minipage}
\end{figure}

As a more exacting test of the various water models and their (M)DCM
representations, the density $\rho$, enthalpy of vaporization $\Delta
H_{\rm vap}$, isobaric heat capacity $C_p$, isothermal compressibility
$\kappa$, and thermal expansion coefficient $\alpha$ were studied as a
function of temperature $T$ between 235.5 K and 350 K. Corresponding
property vs. $T$ profiles are presented alongside reference
experimental data in Figure \ref{fig:propsvst}. In all cases the
TIP$n$P models and their DCM representations agree very
favourably. The same applies to iAMOEBA with the exception of
  $\kappa$ using the DCM representation for which a small shift is
  visible.\\

\noindent
{\bf Liquid density:} The dependence of bulk density $\rho$ on $T$ is
shown in Figure \ref{fig:propsvst}a. In all cases there is close
agreement between the original models and their (M)DCM
representations, and much closer than the agreement between the
different models. The multipolar and MDCM iAMOEBA
descriptions are almost indistinguishable from the experimental curve,
while the DCM description is also very close.\\

\noindent
{\bf Isothermal compressibility and thermal expansion coefficient:}
Plots of isothermal compressibility $\kappa (T)$, and thermal
expansion coefficient $\alpha (T)$ as a function of temperature are
also included in Figures \ref{fig:propsvst}b and c. (M)DCM
representations of TIP$n$P again agree well with reference data, and
more closely than the models agree with one another. The same is true
for iAMOEBA, although some shift is visible in $\kappa$ for the DCM
representation.\\

\noindent
Relative to the experimental results, the additional
detail in the iAMOEBA model, well encapsulated using distributed
charges, affords a consistently accurate performance across the full range
of $T$.\\

\noindent
{\bf Diffusivity:} The self-diffusion coefficient is one of the most
frequently examined transport properties of water in MD studies
\cite{mark2001structure}. It measures the mobility of water molecules
in the H-bonded liquid water network, and is thus taken as an
indication of the accuracy of the water interaction potential. The
simulated $D$, over the range of temperatures studied here, is
presented in Figure \ref{fig:propsvst}d.\\

\noindent
{\bf Enthalpy of vaporization:} The variation of vaporization enthalpy
$\Delta H_{\rm vap}$ with $T$ (Figure \ref{fig:propsvst}e) shows
significant differences between the different models, but in each case
there is good agreement between the DCM implementation and the
existing codes, and between the multipolar, polarizable charge model and
(M)DCM results.\\

\noindent
{\bf Isobaric heat capacity:} For $C_p (T)$, again iAMOEBA and its
(M)DCM representations agree well with one another across a broad
range of temperatures. Slightly more deviation is
visible for the DCM implementation of the TIP5P model. This model
represents a significant outlier, though, in its agreement with
experiment and the remaining models other than at high $T$, which may
lead to increased sensitivity in this property.

\subsection{MDCM for PhF}
Next, the (M)DCM parametrization is extended to solvated systems. As
an example, fluorobenzene (PhF) was chosen. In the past it has been
demonstrated that for halogenated benzenes including detailed
electrostatics is mandatory for quantitative
simulations.\cite{bereau2013,jorgensenHalogens,ClarkSigmaHole}\\

\noindent
The presence of a halogen atom in PhF with a weak ``sigma hole'',
combined with the availability of an existing multipolar
model\cite{bereau2013} make it another suitable choice to evaluate the
impact of replacing multipolar terms with distributed charges. In this
case previously published
experimental\cite{wang_densities,Majer_enthalpies_vap,mobley_solv_energies}
and computed\cite{bereau2013} solvation enthalpies, vaporization
enthalpies of the pure liquid and heat capacities were available for
comparison.\\

\begin{figure}[!ht]
\centering
\begin{minipage}[b]{0.9\textwidth}
\includegraphics[width=0.6\textwidth]{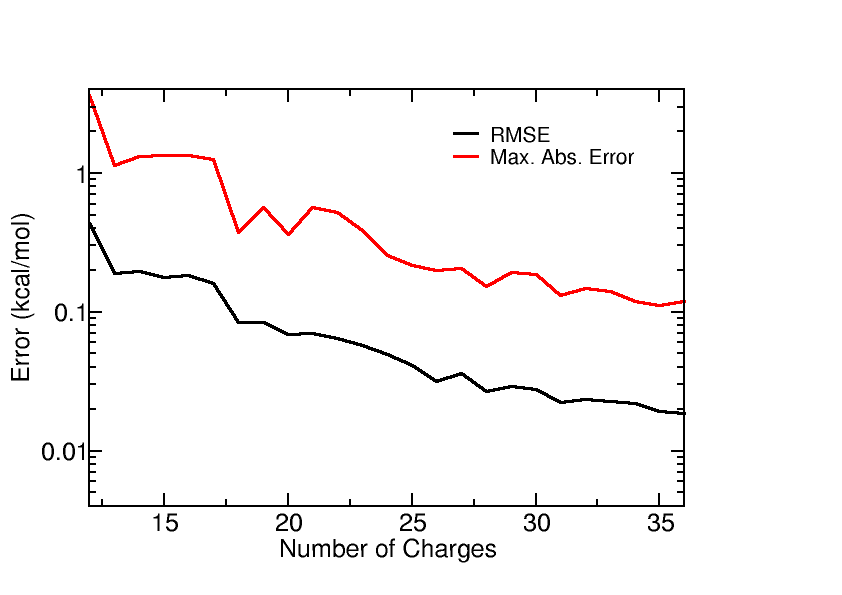}
\includegraphics[width=0.3\textwidth]{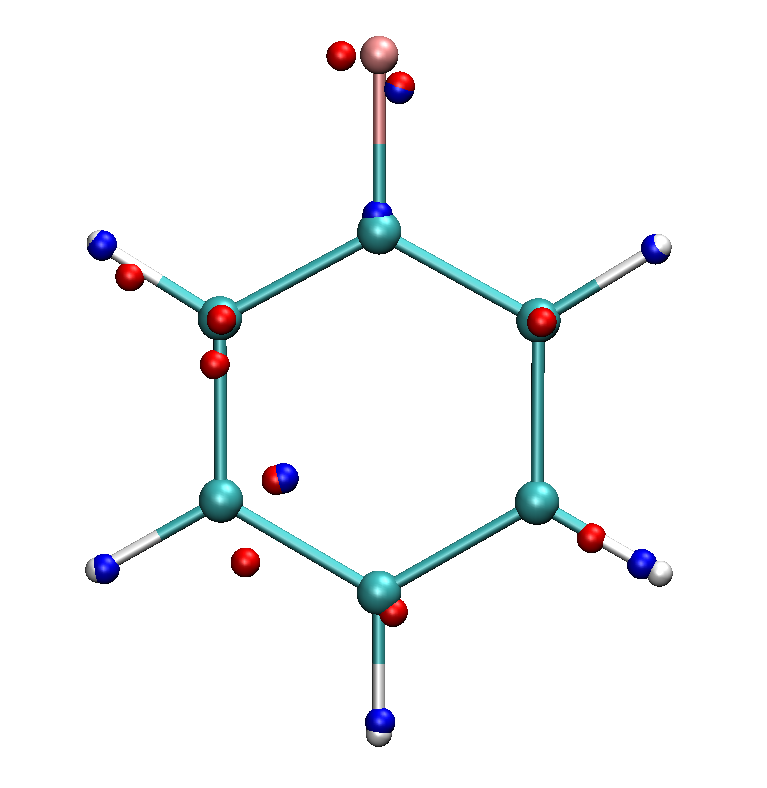}
\caption{Left: RMSE and maximum absolute errors for fits of MDCMs to
  MEP reference data for PhF as a function of increasing number
  of charges. Right: location of the charges (blue positive, red
    negative) for the 22-charge model chosen for condensed phase
  simulations.}
\label{fig:phf-mdcm-fit}
\end{minipage}
\end{figure}

\noindent
As for water, the first step was to obtain suitable MDCMs fitted to a
grid of MEP points generated by the pre-existing multipolar
model. Models were fitted with between 12 and 36 charges, i.e. with an
average of between 1 and 3 charges per atom, see Figure
\ref{fig:phf-mdcm-fit}. As the underlying multipolar model contained
42 non-zero multipolar terms, all charge models again offer a notable
decrease in computational complexity for subsequent simulations. While
the visible noise in Figure \ref{fig:phf-mdcm-fit} with increasing
number of charges in the fit shows that further refinement of the
fitting procedure is possible, for example by increasing the number of
DE fitting generations or the number of fits performed, the generally
systematic improvement is encouraging and offers the possibility to
select an MDCM based on an optimal compromise between computational
cost of simulations due to increased number of charges, and improved
accuracy in the electrostatics.\\

\noindent
It is also encouraging to see that with 13 charges, i.e. an average of
1.1 charges per atom, the RMSE has already dropped to 0.19 kcal/mol
with a maximum absolute error across the grid of 1.14 kcal/mol. With
18 charges the RMSE is 0.08 kcal/mol and the maximum absolute error is
0.38 kcal/mol and with an average of 3 charges per atom the accuracy
reaches an RMSE of 0.02 kcal/mol and the maximum absolute error is
0.12 kcal/mol.\\

\subsection{Thermodynamic Properties of PhF}
\label{section:PhFProps}
After fitting MDCMs for PhF, the performance of the models in
condensed phase simulations with respect to the performance of the
original multipolar model was investigated. For this, the 22-charge
MDCM was selected as it offered good accuracy (an RMSE of 0.06
kcal/mol across the electrostatic grid used for fitting and a maximum
absolute error of 0.53 kcal/mol) at modest computational cost with
fewer than 2 charges on average per atom (Figure
\ref{fig:phf-mdcm-fit}). Again, as no symmetry constraints were
applied the charge positions are spatially asymmetric, but the close
agreement in MEP with the symmetric underlying multipolar model
demonstrates that the symmetry of the MEP is conserved.\\

\noindent
The thermodynamic properties chosen for comparison with available
experimental data were the density of pure liquid PhF, the
vaporization enthalpy of liquid PhF and the solvation free energy of a
single PhF molecule in liquid water. As simulation conditions differed
slightly from those used to fit the multipolar model\cite{bereau2013}
a scaling factor of 1.1 was applied to the Lennard-Jones `$\sigma$'
and `$\epsilon$' parameters of heavy atoms to recover the original
performance of the multipolar model in describing these
properties. The same scaling factor was applied to L-J parameters in
MDCM simulations to allow direct comparison.\\

\noindent
Results averaged over 8 independent simulations are shown in Table
\ref{tab:dg_phf}. While both models agree quite well with experiment,
there is again very good agreement between the multipolar and MDCM
condensed-phase simulation results. That this agreement is again
possible without refitting any other force-field terms and by fitting
only to MEP grid-data highlights the equivalent performance of a
modest number of distributed charges to a full multipole-expansion
truncated at quadrupole when describing the dynamics of these
systems.\\

\begin{table}[!ht]
\caption{Calculated free energy $\Delta G$ (kcal/mol) of solvation of
  a single molecule of PhF in a TIP3P solvent box (averaged over 8
  thermodynamic integration runs, with contributions from
  electrostatic and vdW terms), $\Delta H$ of vaporization (kcal/mol)
  and density at 300 K of pure liquid PhF. Results using an existing
  multipolar model for PhF are compared with results from a fitted
  MDCM with 22 charges. The standard deviation for each computed value
  is reported next to the number.}
\begin{center}
\begin{tabular}{lccccc}
\hline
\hline
\multicolumn{1}{ l }{}               &
\multicolumn{1}{ c }{$\rho$}         &
\multicolumn{1}{ c }{$\Delta H_{\rm vap}$} &
\multicolumn{1}{ c }{$\Delta G_{\rm solv,vdw}^c$} &
\multicolumn{1}{ c }{$\Delta G_{\rm solv,elec}$}  &
\multicolumn{1}{ c }{$\Delta G_{\rm solv,tot}$}         \\
\hline
Multipolar$^a$   & 0.90{$\pm 0.01$}      & 9.37{$\pm 0.05$}      & 2.21{$\pm 0.13$} $\dagger$ & --2.32 {$\pm 0.04$}& --0.10 {$\pm 0.14$}\\
MDCM             & 0.90{$\pm 0.01$}      & 9.39{$\pm 0.06$}      & 2.21{$\pm 0.13$} $\dagger$ & --2.92 {$\pm 0.13$}& --0.71 {$\pm 0.21$}\\
Exp.             & 1.02$^b$ & 8.26$^c$ & -              & -     & --0.80$^d$ \\
\hline
\hline
\end{tabular}\\
$^a$ Ref.\cite{bereau2013} ; $^b$ Ref.\cite{wang_densities} ; $^c$
Ref.\cite{Majer_enthalpies_vap} ; $^d$ Ref.\cite{mobley_solv_energies}
.\\ $\dagger$ The same 8 trajectories were used to evaluate Multipolar
and MDCM vdW solvation energy contributions as solute electrostatics are zeroed in these calculations.
\end{center}
\label{tab:dg_phf}
\end{table}

\subsection{Computational Efficiency}
 While the results so far have demonstrated the equivalence
in accuracy of distributed point charge and multipolar models in
polarizable (iAMOEBA) and non-polarizable (PhF) MD simulations, any
advantage over multipolar methods depends additionally on relative
computational cost and ease of implementation into general-purpose MD
codes. The second point has been demonstrated by introducing
  (M)DCMs for H$_{2}$O into OpenMM by using only existing
  functionality. This suggests that (M)DCM can also be used in any
  other MD simulation code that provide appropriate functionality
  without undue additional technical effort.\\

\noindent
The first point is more difficult to quantify in an objective manner
as it depends on factors including hardware (CPU or GPU
specifications, hardware optimization for important terms such as
inverse square roots), compiler details, algorithmic efficiency, size
of the simulated system, treatment of electrostatics (PME, cut-off
distances), numerical precision and others. A rigorous treatment falls
beyond the scope of the current work, but further to a brief analysis
presented previously,\cite{devereux2014novel} analyses and timings for
the simulations discussed here are given in section 10 of the SI. The observation is that MDCMs outperform MTPs, ranging from
  $\sim 20$ \% for water to a factor of three for PhF. However,
  additional analysis for other molecules and types of simulations is
  required to yield conclusive results.

\section{Conclusions}
 (M)DCM has been improved in accuracy, generalized and extended to
include polarizability and applied to condensed phase simulations. Its
performance to capture multipolar interactions within a single point
charge-based implementation has been demonstrated by comparing various
condensed phase properties of water and solvated PhF with established
multipolar representations. Specifically, the distributed charge
framework (including distribution of torques, local axis system
description, definition of cut-offs) maintains the original bulk
simulation properties for traditional force field models such as
TIP3P, models with one or more off-centered charges such as TIP4P and
TIP5P, and multipolar models such as the recently developed iAMOEBA
and a multipolar description of PhF. Although there are different
possible ways to define axis systems (Figure \ref{fig:lra}) and to
distribute the torques, the close agreement and consistency of the
results from the MD simulations for eight condensed phase properties
using the different models indicates that these choices are not
critical to describing the dynamics of the system. This is not
self-evident and a gratifying aspect of the present work.\\

\noindent
The use of point charges in place of atom-centered multipoles removes
the need for computationally expensive higher-order multipole terms to
yield accurate and efficient dynamics.  Furthermore, machine-learning
reduces the number of charges necessary to a minimal set and the
method has been combined with isotropic polarization and integrated
with familiar tools such as barostats and thermodynamic integration
routines to enable straightforward use of the models under widely
encountered condensed-phase simulation conditions. The compatibility
of the approach with an unmodified release of OpenMM and with all of
the force field models tested here shows that distributed charges are
a versatile tool for the development of next-generation force fields
and multi-level approaches, and also offer an alternative to explicit
multipolar terms of existing force fields. Beyond the simulation
packages presented already, widely used codes such as
Amber\cite{case2005amber} also offer functionality for off-centered
charges that could be combined with (M)DCM without further
modification or prohibitive computational cost.\\

\noindent
While the current work focused on directly comparing simulation
results from (M)DCM with multipolar representations {\it without
  refitting} remaining parameters for direct comparison of the
implementations, a generic (M)DCM force field, fitted to condensed
phase experiments offers additional opportunities for quantitative
simulations. This can be envisaged within the framework of an existing
fitting environment\cite{fittingwizard2016} and will be of particular
interest when balancing accuracy (i.e. the number of charges per atom)
and speed for specific applications. As was shown here it is possible
by increasing the number of charges used in fitting to systematically
improve the accuracy of the electrostatic interactions, allowing fine
tuning of the computational cost of adding more charges and a
corresponding improvement in the description of the electrostatic
interaction. In this way, the present work opens up the possibility
for custom-made MDCM force fields with calibrated accuracy to
encapsulate the physics of a given application.\\

\section{Supplementary Material}
See supplementary material for an illustration of the complexity of
multipolar versus distributed charge interaction terms, for parameter
files with (M)DCM charge positions and magnitudes, for sample input
files used to run (M)DCM in OpenMM, for radial distribution functions
from different water model simulations, for a discussion of the
relationship between errors in the electrostatic interaction energy
and fitting errors in the MEP, for an analysis of the impact of broken symmetry in MDCMs on interaction energies, and for a discussion of relative computational costs of MDCMs and multipole moments. The code used to fit the MDCMs presented here is available freely at https://github.com/MMunibas/MDCM.

\begin{acknowledgement}
The authors acknowledge financial support from the Swiss National 
Foundation for Research for their support within the NCCR 
MUST program and project 200021-117810.
\end{acknowledgement}

\bibliography{literature}

\newpage
\clearpage

\begin{figure}[!ht]
\centering
\begin{minipage}[b]{0.9\textwidth}
\includegraphics[width=3.5in]{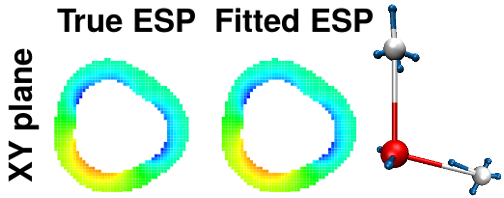}
\caption{TOC graphic}
\end{minipage}
\end{figure}

\end{document}


\maketitle
\pagebreak
\tableofcontents

\pagebreak

\section{Reduction in complexity using distributed charges}
\label{sec:reduction}
The main benefit of using distributed charges over a truncated
multipole expansion is that the electrostatic interaction energy
between two multipolar atoms can be evaluated using a homogeneous set
of charge-charge interaction terms in place of the heterogeneous terms
between multipole moments of different ranks. I.e. all energy terms
are of the form:

\begin{align}
V_{\rm ab} &= \frac{q_{\rm a}q_{{\rm b}}}{R_{\rm ab}} \label{Eq:v00}
\end{align}

\noindent
where $R_{\rm ab}$ is the distance between atoms $a$ and $b$. By contrast,
multipolar interaction terms are heterogeneous and especially higher
rank terms are substantially more complex. The following example
describes the interaction between multipole moments $Q_{l,m}$ on atoms $a$
and $d$\cite{stonebook}:

\begin{align}
V_{\rm Q_{00,{\rm a}}Q_{00,{\rm d}}} &= \frac{Q_{\rm 00,a}Q_{00,{\rm d}}}{r_{\rm ad}} \label{Eq:v0000} \\
V_{\rm Q_{00,{\rm a}}Q_{1m,{\rm d}}} &= \frac{Q_{\rm 00,a}Q_{1m,{\rm d}}r_{\alpha}^{\rm a}}{r_{\rm ad}^{2}} \label{Eq:v001m} \\
V_{\rm Q_{1m,{\rm a}}Q_{1m},{\rm d}} &= \frac{Q_{1m,{\rm a}}Q_{1m,{\rm d}}(3r_{\alpha}^{\rm a}r_{\beta}^{\rm d}+c_{\alpha,\beta})}{r_{\rm ad}^{3}} \label{Eq:v1m1m} \\
V_{\rm Q_{20,{\rm a}}Q_{00},{\rm d}} &= \frac{Q_{20,{\rm a}}Q_{00,{\rm d}}(3r_{z}^{\rm a 2}-1)}{2r_{\rm ad}^{3}} \label{Eq:v201m} \\
V_{\rm Q_{22c,{\rm a}}Q_{00},{\rm d}} &= \frac{\sqrt{3}Q_{22c,{\rm a}}Q_{00,{\rm d}}(r_{x}^{\rm a  2}-r_{y}^{\rm a  2})}{2r_{\rm ad}^{3}} \label{Eq:v22c1m} \\
V_{\rm Q_{20,{\rm a}}Q_{1m,{\rm d}}} &= \frac{Q_{20,{\rm a}}Q_{1m,{\rm d}}(15r_{z}^{\rm a 2} r_{\beta}^{\rm d}+6 c_{zA,\beta}r_{z}^{\rm a}-3 r_{\beta}^{\rm d})}{2 r_{\rm ad}^{4}} \label{Eq:v1m20} \\
V_{\rm Q_{22c,{\rm a}}Q_{1m,{\rm d}}} &= \frac{\sqrt{3}Q_{22c,{\rm a}}Q_{1m,{\rm d}}(5 r_{\beta}^{\rm d}(r_{x}^{\rm a 2} - r_{y}^{\rm a 2})+2 c_{xA,\beta}r_{x}^{\rm a}-2 c_{yA,\beta}r_{y}^{\rm a})}{2 r_{\rm ad}^{4}} \label{Eq:v1m22c} \\
V_{\rm Q_{20,{\rm a}}Q_{20,{\rm d}}} &= \frac{3 Q_{20,{\rm a}}Q_{20,{\rm d}}(35 r_{z}^{\rm a 2}r_{z}^{\rm d 2} - 5 r_{z}^{\rm a 2}-5 r_{z}^{\rm d 2}+20 r_{z}^{\rm a} r_{z}^{\rm d} c_{zA,zD}+2 c_{zA,zD}^{2}+1)}{4 r_{\rm ad}^{5}} \label{Eq:v2020} \\
V_{\rm Q_{20,{\rm a}}Q_{22c,{\rm d}}} &= \frac{\sqrt{3} Q_{20,{\rm a}}Q_{22c,{\rm d}}(35 r_{z}^{\rm a 2}r_{x}^{\rm d}r_{y}^{\rm d} - 5 r_{x}^{\rm d}r_{y}^{\rm d}+10 r_{z}^{\rm a}r_{x}^{\rm d}c_{zA,yD}+10 r_{z}^{\rm a} r_{y}^{\rm d} c_{zA,xD}+2 c_{zA,xD} c_{zA,yD})}{2 r_{\rm ad}^{5}} \label{Eq:v2022c} \\
V_{\rm Q_{22c,{\rm a}}Q_{22c,{\rm d}}} &= \frac{1}{4} Q_{22c,{\rm a}}Q_{22c,{\rm d}}(35 r_{x}^{\rm a 2}r_{x}^{\rm d 2}-35r_{x}^{\rm a 2}r_{y}^{\rm d 2} r_{y}^{\rm d}-35r_{y}^{\rm a 2}r_{x}^{\rm d 2}+35r_{y}^{\rm a 2}r_{y}^{\rm d 2} r_{y}^{\rm d} r_{y}^{\rm d} \notag \\
 &\qquad+20 r_{x}^{\rm a}r_{x}^{\rm d}c_{xA,xD}-20 r_{x}^{\rm a}r_{y}^{\rm d}c_{xA,yD} -20 r_{y}^{\rm a}r_{x}^{\rm d}c_{yA,xD} +20 r_{y}^{\rm a}r_{y}^{\rm d}c_{yA,yD} \notag \\
 &\qquad + 2 c_{xA,xD}^{2}-2 c_{xA,yD}^{2}-2 c_{yA,xD}^{2}+2 c_{yA,yD}^{2}) r_{\rm ad}^{-5} \label{Eq:v22c22c}
\end{align}

\noindent Here the multipole moment $Q_{1m}$ is the dipole moment component $m$ ($l=1$). When $m=0,1,2$, $\alpha=z,x,y$ respectively. $r_{\rm ad}$ is the internuclear separation of atoms \texttt{a} and \texttt{d}, $r_{\alpha}^{\rm a}$ is defined as the scalar product $\hat{\mathbf{e}}_{\alpha}^{\rm a}\cdot\hat{\mathbf{e}}_{\rm ad}$ where $\hat{\mathbf{e}}_{\alpha}^{\rm a}$ is a unit vector along the $\alpha$ axis of atom \texttt{a} ($\alpha=x,y,z$). $\hat{\mathbf{e}}_{\rm ad}$ is a unit vector in the direction from \texttt{a} to \texttt{d}. $\beta=x,y,z$ for atom \texttt{d}, $c_{\alpha,\beta}$ is the scalar product $\hat{\mathbf{e}}_{\alpha}^{\rm a}\cdot\hat{\mathbf{e}}_{\beta}^{\rm d}$.
 
In order to run molecular
dynamics simulations the gradients of each of these terms must also be
evaluated for every nonbonded atom pair at each simulation time step.

\newpage
\clearpage


\section{(M)DCM Charge Models}

\begin{table}[!ht]
\caption{Octahedrally arranged DCM charges (a.u.) and their
  coordinates (\AA) for H$_2$O used in iAMOEBA model; atomic
  polarizabilities $\alpha$ in \AA$^3$. Coordinates are relative to
  local atomic axes.\cite{devereux2014novel}}
\begin{center}
\begin{tabular}{lcccccccccc}
\hline
\hline
        && $X$        && $Y$        && $Z$      && $q$    && $\alpha$         \\
\hline
{\bf H} &&                 &&                    &&                   &&                    && 0.50484            \\
$q_{1}$ && -0.0956010671   &&    0.0000000000   &&    0.0090474640   &&    1.6134817638   &&                    \\
$q_{2}$ &&  0.0000000000   &&    0.0959947647   &&    0.0000000000   &&    0.2524281689   &&                    \\
$q_{3}$ &&  0.0956010671   &&    0.0000000000   &&   -0.0090474640   &&    2.0482594186   &&                    \\
$q_{4}$ && -0.0086850918   &&    0.0000000000   &&   -0.0995898760   &&   -2.2882340867   &&                    \\
$q_{5}$ &&  0.0000000000   &&   -0.0959947647   &&    0.0000000000   &&    0.2524281689   &&                    \\
$q_{6}$ && 0.0086850918   &&    0.0000000000   &&    0.0995898760   &&   -1.5813534334   &&                    \\
\hline                                                                                                            
{\bf O} &&                 &&                    &&                   &&                    && 0.80636            \\
$q_{1}$ && -0.0575896814   &&    0.0000000000   &&   -0.0800056071   &&    2.0122163666   &&                    \\
$q_{2}$ &&  0.0000000000   &&   -0.0959947647   &&    0.0000000000   &&   -3.1087375276   &&                    \\
$q_{3}$ &&  0.0575896814   &&    0.0000000000   &&    0.0800056071   &&    2.0122163666   &&                    \\
$q_{4}$ && -0.0768011943   &&    0.0000000000   &&    0.0599925231   &&    1.0336186790   &&                    \\
$q_{5}$ &&  0.0000000000   &&    0.0959947647   &&    0.0000000000   &&   -3.1087375276   &&                    \\
$q_{6}$ && 0.0768011943   &&    0.0000000000   &&   -0.0599925231   &&    0.5654036430   &&                    \\
\hline                                                                                                            
{\bf H} &&                 &&                    &&                   &&                    && 0.50484            \\
$q_{1}$ &&  0.0956010671   &&    0.0000000000    &&    0.0090474640   &&     1.6134817638   &&                    \\
$q_{2}$ &&  0.0000000000   &&   -0.0959947647    &&    0.0000000000   &&     0.2524281689   &&                    \\
$q_{3}$ && -0.0956010671   &&    0.0000000000    &&   -0.0090474640   &&     2.0482594186   &&                    \\
$q_{4}$ &&  0.0086850918   &&    0.0000000000    &&   -0.0995898760   &&    -2.2882340867   &&                    \\
$q_{5}$ &&  0.0000000000   &&    0.0959947647    &&    0.0000000000   &&     0.2524281689   &&                    \\
$q_{6}$ &&  -0.0086850918   &&    0.0000000000    &&    0.0995898760   &&    -1.5813534334   &&                    \\
\hline
\hline
\end{tabular}
\end{center}
\label{tab:dcmiampar}
\end{table}

\newpage
\clearpage


\begin{table}[!ht]
\caption{MDCM charges (a.u.) and their positions (\AA) for 10-charge
  H$_2$O model used for iAMOEBA; atomic polarizabilities $\alpha$ in
  \AA$^3$. Coordinates are relative to local atomic
  axes.\cite{devereux2014novel} }
\begin{center}
\begin{tabular}{lcccccccccc}
\hline
\hline
        && $X$        && $Y$        && $Z$      && $q$    && $\alpha$         \\
\hline
{\bf H} &&                 &&                    &&                   &&                    && 0.50484            \\
$q_{1}$ &&  0.0209144768   &&    0.0000004420   &&    0.3473443817   &&   -0.1066619461   &&                    \\
$q_{2}$ && -0.0452396324   &&    0.0000004981   &&   -0.1651346312   &&   -0.9725811260   &&                    \\
$q_{3}$ && -0.1905254480   &&    0.0000021077   &&   -0.0528319716   &&    0.5997075639   &&                    \\
$q_{4}$ &&  0.1496075438   &&   -0.0018910650   &&   -0.0565055370   &&    0.8289103944   &&                    \\
\hline                                                                                                            
{\bf O} &&                 &&                    &&                   &&                    && 0.80636            \\
$q_{1}$ &&  0.1009343674   &&    0.4067419560   &&   -0.0294610237   &&   -0.2369290661   &&                    \\
$q_{2}$ &&  0.0881957994   &&   -0.3169912077   &&   -0.0203134193   &&   -0.3142376636   &&                    \\
\hline                                                                                                            
{\bf H} &&                 &&                    &&                   &&                    && 0.50484            \\
$q_{1}$ &&  0.1659300202   &&   -0.0008225336   &&    0.0275557737   &&    0.5465716486   &&                    \\
$q_{2}$ && -0.1837974669   &&   -0.0020877351   &&   -0.0477859191   &&    0.6204782536   &&                    \\
$q_{3}$ &&  0.0519918506   &&   -0.0000018568   &&   -0.3764240427   &&   -0.6244927174   &&                    \\
$q_{4}$ &&  0.0355028536   &&   -0.0000020937   &&    0.2009935462   &&   -0.3407653412   &&                    \\
\hline
\hline
\end{tabular}
\end{center}
\label{tab:dcmiampar}
\end{table}

\newpage
\clearpage


\section{iAMOEBA DCM OpenMM force field definition file}
The following "XML" format force field definition file defines the
iAMOEBA/DCM implementation for use in OpenMM. While dummy atom types
could potentially be consolidated (all dummy atoms share the same mass
and VDW, sites belonging to the same atom-type also have the same
polarizability), the exhaustive approach taken here is convenient for
development purposes.

\lstset{language=XML}
\begin{lstlisting}
<ForceField>
 <AtomTypes>
  <Type name="380" class="73" element="O" mass="15.999"/>
  <Type name="381" class="74" element="H" mass="1.008"/>
  <Type name="382" class="75" mass="0"/>
  <Type name="383" class="76" mass="0"/>
  <Type name="384" class="77" mass="0"/>
  <Type name="385" class="78" mass="0"/>
  <Type name="386" class="79" mass="0"/>
  <Type name="387" class="80" mass="0"/>
  <Type name="388" class="81" mass="0"/>
  <Type name="389" class="82" mass="0"/>
  <Type name="390" class="83" mass="0"/>
  <Type name="391" class="84" mass="0"/>
  <Type name="392" class="85" mass="0"/>
  <Type name="393" class="86" mass="0"/>
  <Type name="394" class="87" mass="0"/>
  <Type name="395" class="88" mass="0"/>
  <Type name="396" class="89" mass="0"/>
  <Type name="397" class="90" mass="0"/>
  <Type name="398" class="91" mass="0"/>
  <Type name="399" class="92" mass="0"/>
 </AtomTypes>
 <Residues>
  <Residue name="HOH">
   <Atom name="O" type="380"/>
   <Atom name="H1" type="381"/>
   <Atom name="H2" type="381"/>
   <Atom name="M1" type="382"/>
   <Atom name="M2" type="383"/>
   <Atom name="M3" type="384"/>
   <Atom name="M4" type="385"/>
   <Atom name="M5" type="386"/>
   <Atom name="M6" type="387"/>
   <Atom name="M7" type="388"/>
   <Atom name="M8" type="389"/>
   <Atom name="M9" type="390"/>
   <Atom name="M10" type="391"/>
   <Atom name="M11" type="392"/>
   <Atom name="M12" type="393"/>
   <Atom name="M13" type="394"/>
   <Atom name="M14" type="395"/>
   <Atom name="M15" type="396"/>
   <Atom name="M16" type="397"/>
   <Atom name="M17" type="398"/>
   <Atom name="M18" type="399"/>
   <Bond from="0" to="2"/>
   <Bond from="0" to="1"/>

<!-- H1-atom charges and positions -->
   <VirtualSite type="localCoords" index="3" atom1="0" atom2="1" atom3="2"  wo1="0.0" wo2="1.0" wo3="0.0"  wx1="-1.0" wx2="1.0"  wx3="0.0" wy1="-1.0" wy2="0.0" wy3="1.0"  p1=" 0.000905"  p2=" 0.009551"  p3=" 0.000000" />
   <VirtualSite type="localCoords" index="4" atom1="0" atom2="1" atom3="2"  wo1="0.0" wo2="1.0" wo3="0.0"  wx1="-1.0" wx2="1.0"  wx3="0.0" wy1="-1.0" wy2="0.0" wy3="1.0"  p1=" 0.000000"  p2="-0.000000"  p3=" 0.009591" />
   <VirtualSite type="localCoords" index="5" atom1="0" atom2="1" atom3="2"  wo1="0.0" wo2="1.0" wo3="0.0"  wx1="-1.0" wx2="1.0"  wx3="0.0" wy1="-1.0" wy2="0.0" wy3="1.0"  p1="-0.000905"  p2="-0.009551"  p3=" 0.000000" />
   <VirtualSite type="localCoords" index="6" atom1="0" atom2="1" atom3="2"  wo1="0.0" wo2="1.0" wo3="0.0"  wx1="-1.0" wx2="1.0"  wx3="0.0" wy1="-1.0" wy2="0.0" wy3="1.0"  p1="-0.009959"  p2=" 0.000868"  p3=" 0.000000" />
   <VirtualSite type="localCoords" index="7" atom1="0" atom2="1" atom3="2"  wo1="0.0" wo2="1.0" wo3="0.0"  wx1="-1.0" wx2="1.0"  wx3="0.0" wy1="-1.0" wy2="0.0" wy3="1.0"  p1=" 0.000000"  p2="-0.000000"  p3="-0.009591" />
   <VirtualSite type="localCoords" index="8" atom1="0" atom2="1" atom3="2"  wo1="0.0" wo2="1.0" wo3="0.0"  wx1="-1.0" wx2="1.0"  wx3="0.0" wy1="-1.0" wy2="0.0" wy3="1.0"  p1=" 0.009959"  p2="-0.000868"  p3=" 0.000000" />
<!-- O-atom charges and positions -->
   <VirtualSite type="localCoords" index="9" atom1="0" atom2="1" atom3="2"  wo1="1.0" wo2="0.0" wo3="0.0"  wx1="-1.0" wx2="1.0"  wx3="0.0" wy1="-1.0" wy2="0.0" wy3="1.0"  p1="-0.008010"  p2=" 0.005742"  p3=" 0.000000" />
   <VirtualSite type="localCoords" index="10" atom1="0" atom2="1" atom3="2"  wo1="1.0" wo2="0.0" wo3="0.0"  wx1="-1.0" wx2="1.0"  wx3="0.0" wy1="-1.0" wy2="0.0" wy3="1.0"  p1=" 0.000000"  p2="-0.000000"  p3="-0.009591" />
   <VirtualSite type="localCoords" index="11" atom1="0" atom2="1" atom3="2"  wo1="1.0" wo2="0.0" wo3="0.0"  wx1="-1.0" wx2="1.0"  wx3="0.0" wy1="-1.0" wy2="0.0" wy3="1.0"  p1=" 0.008010"  p2="-0.005742"  p3=" 0.000000" />
   <VirtualSite type="localCoords" index="12" atom1="0" atom2="1" atom3="2"  wo1="1.0" wo2="0.0" wo3="0.0"  wx1="-1.0" wx2="1.0"  wx3="0.0" wy1="-1.0" wy2="0.0" wy3="1.0"  p1=" 0.005987"  p2=" 0.007682"  p3=" 0.000000" />
   <VirtualSite type="localCoords" index="13" atom1="0" atom2="1" atom3="2"  wo1="1.0" wo2="0.0" wo3="0.0"  wx1="-1.0" wx2="1.0"  wx3="0.0" wy1="-1.0" wy2="0.0" wy3="1.0"  p1=" 0.000000"  p2="-0.000000"  p3=" 0.009591" />
   <VirtualSite type="localCoords" index="14" atom1="0" atom2="1" atom3="2"  wo1="1.0" wo2="0.0" wo3="0.0"  wx1="-1.0" wx2="1.0"  wx3="0.0" wy1="-1.0" wy2="0.0" wy3="1.0"  p1="-0.005987"  p2="-0.007682"  p3=" 0.000000" />
<!-- H2-atom charges and positions -->
   <VirtualSite type="localCoords" index="15" atom1="0" atom2="1" atom3="2"  wo1="0.0" wo2="0.0" wo3="1.0"  wx1="-1.0" wx2="0.0"  wx3="1.0" wy1="1.0" wy2="-1.0" wy3="0.0"  p1=" 0.000905"  p2="-0.009551"  p3=" 0.000000" />
   <VirtualSite type="localCoords" index="16" atom1="0" atom2="1" atom3="2"  wo1="0.0" wo2="0.0" wo3="1.0"  wx1="-1.0" wx2="0.0"  wx3="1.0" wy1="1.0" wy2="-1.0" wy3="0.0"  p1=" 0.000000"  p2="-0.000000"  p3="-0.009591" />
   <VirtualSite type="localCoords" index="17" atom1="0" atom2="1" atom3="2"  wo1="0.0" wo2="0.0" wo3="1.0"  wx1="-1.0" wx2="0.0"  wx3="1.0" wy1="1.0" wy2="-1.0" wy3="0.0"  p1="-0.000905"  p2=" 0.009551"  p3=" 0.000000" />
   <VirtualSite type="localCoords" index="18" atom1="0" atom2="1" atom3="2"  wo1="0.0" wo2="0.0" wo3="1.0"  wx1="-1.0" wx2="0.0"  wx3="1.0" wy1="1.0" wy2="-1.0" wy3="0.0"  p1="-0.009959"  p2="-0.000868"  p3=" 0.000000" />
   <VirtualSite type="localCoords" index="19" atom1="0" atom2="1" atom3="2"  wo1="0.0" wo2="0.0" wo3="1.0"  wx1="-1.0" wx2="0.0"  wx3="1.0" wy1="1.0" wy2="-1.0" wy3="0.0"  p1=" 0.000000"  p2="-0.000000"  p3=" 0.009591" />
   <VirtualSite type="localCoords" index="20" atom1="0" atom2="1" atom3="2"  wo1="0.0" wo2="0.0" wo3="1.0"  wx1="-1.0" wx2="0.0"  wx3="1.0" wy1="1.0" wy2="-1.0" wy3="0.0"  p1=" 0.009959"  p2=" 0.000868"  p3=" 0.000000" />
 </Residue>
 </Residues>
   <AmoebaBondForce bond-cubic="-25.5" bond-quartic="379.3125">
   <Bond class1="73" class2="74" length="9.584047e-02" k="2.3331232e+05"/>
 </AmoebaBondForce>
  <AmoebaAngleForce angle-cubic="-0.014" angle-quartic="5.6e-05" angle-pentic="-7e-07" angle-sextic="2.2e-08">
   <Angle class1="74" class2="73" class3="74" k="6.359379296918e-02" angle1="1.064826e+02"/>
 </AmoebaAngleForce>
  <AmoebaOutOfPlaneBendForce type="ALLINGER" opbend-cubic="-0.014" opbend-quartic="5.6e-05" opbend-pentic="-7e-07" opbend-sextic="2.2e-08">
    <!-- LPW: Mark's force field parsing code requires AmoebaOutOfPlaneBendForce in order to read AmoebaAngleForce, even if the clause is empty -->
 </AmoebaOutOfPlaneBendForce>
  <AmoebaVdwForce type="BUFFERED-14-7" radiusrule="CUBIC-MEAN" radiustype="R-MIN" radiussize="DIAMETER" epsilonrule="HHG" vdw-13-scale="0.0" vdw-14-scale="1.0" vdw-15-scale="1.0">
   <Vdw class="73" sigma="3.645297e-01" epsilon="8.2348e-01" reduction="1.0"/>
   <Vdw class="74" sigma="0.0" epsilon="0.0" reduction="1.0"/>
   <Vdw class="75" sigma="0.0" epsilon="0.0" reduction="1.0"/>
   <Vdw class="76" sigma="0.0" epsilon="0.0" reduction="1.0"/>
   <Vdw class="77" sigma="0.0" epsilon="0.0" reduction="1.0"/>
   <Vdw class="78" sigma="0.0" epsilon="0.0" reduction="1.0"/>
   <Vdw class="79" sigma="0.0" epsilon="0.0" reduction="1.0"/>
   <Vdw class="80" sigma="0.0" epsilon="0.0" reduction="1.0"/>
   <Vdw class="81" sigma="0.0" epsilon="0.0" reduction="1.0"/>
   <Vdw class="82" sigma="0.0" epsilon="0.0" reduction="1.0"/>
   <Vdw class="83" sigma="0.0" epsilon="0.0" reduction="1.0"/>
   <Vdw class="84" sigma="0.0" epsilon="0.0" reduction="1.0"/>
   <Vdw class="85" sigma="0.0" epsilon="0.0" reduction="1.0"/>
   <Vdw class="86" sigma="0.0" epsilon="0.0" reduction="1.0"/>
   <Vdw class="87" sigma="0.0" epsilon="0.0" reduction="1.0"/>
   <Vdw class="88" sigma="0.0" epsilon="0.0" reduction="1.0"/>
   <Vdw class="89" sigma="0.0" epsilon="0.0" reduction="1.0"/>
   <Vdw class="90" sigma="0.0" epsilon="0.0" reduction="1.0"/>
   <Vdw class="91" sigma="0.0" epsilon="0.0" reduction="1.0"/>
   <Vdw class="92" sigma="0.0" epsilon="0.0" reduction="1.0"/>
 </AmoebaVdwForce>
  <AmoebaMultipoleForce direct11Scale="0.0" direct12Scale="1.0" direct13Scale="1.0" direct14Scale="1.0" mpole12Scale="0.0" mpole13Scale="0.0" mpole14Scale="0.4" mpole15Scale="0.8" mutual11Scale="1.0" mutual12Scale="1.0" mutual13Scale="1.0" mutual14Scale="1.0" polar12Scale="0.0" polar13Scale="0.0" polar14Intra="0.5" polar14Scale="1.0" polar15Scale="1.0">
   <Multipole type="380" c0="0.0" d1="0.0" d2="0.0" d3="0.0" q11="0.0" q21="0.0" q22="0.0" q31="0.0" q32="0.0" q33="0.0"/>
   <Multipole type="381" c0="0.0" d1="0.0" d2="0.0" d3="0.0" q11="0.0" q21="0.0" q22="0.0" q31="0.0" q32="0.0" q33="0.0"/>
   <Multipole type="382" c0="1.613482" d1="0.0" d2="0.0" d3="0.0" q11="0.0" q21="0.0" q22="0.0" q31="0.0" q32="0.0" q33="0.0"/>
   <Multipole type="383" c0="0.252428" d1="0.0" d2="0.0" d3="0.0" q11="0.0" q21="0.0" q22="0.0" q31="0.0" q32="0.0" q33="0.0"/>
   <Multipole type="384" c0="2.048259" d1="0.0" d2="0.0" d3="0.0" q11="0.0" q21="0.0" q22="0.0" q31="0.0" q32="0.0" q33="0.0"/>
   <Multipole type="385" c0="-2.288234" d1="0.0" d2="0.0" d3="0.0" q11="0.0" q21="0.0" q22="0.0" q31="0.0" q32="0.0" q33="0.0"/>
   <Multipole type="386" c0="0.252428" d1="0.0" d2="0.0" d3="0.0" q11="0.0" q21="0.0" q22="0.0" q31="0.0" q32="0.0" q33="0.0"/>
   <Multipole type="387" c0="-1.581353" d1="0.0" d2="0.0" d3="0.0" q11="0.0" q21="0.0" q22="0.0" q31="0.0" q32="0.0" q33="0.0"/>
   <Multipole type="388" c0="2.012216" d1="0.0" d2="0.0" d3="0.0" q11="0.0" q21="0.0" q22="0.0" q31="0.0" q32="0.0" q33="0.0"/>
   <Multipole type="389" c0="-3.108738" d1="0.0" d2="0.0" d3="0.0" q11="0.0" q21="0.0" q22="0.0" q31="0.0" q32="0.0" q33="0.0"/>
   <Multipole type="390" c0="2.012216" d1="0.0" d2="0.0" d3="0.0" q11="0.0" q21="0.0" q22="0.0" q31="0.0" q32="0.0" q33="0.0"/>
   <Multipole type="391" c0="1.033619" d1="0.0" d2="0.0" d3="0.0" q11="0.0" q21="0.0" q22="0.0" q31="0.0" q32="0.0" q33="0.0"/>
   <Multipole type="392" c0="-3.108738" d1="0.0" d2="0.0" d3="0.0" q11="0.0" q21="0.0" q22="0.0" q31="0.0" q32="0.0" q33="0.0"/>
   <Multipole type="393" c0="0.565404" d1="0.0" d2="0.0" d3="0.0" q11="0.0" q21="0.0" q22="0.0" q31="0.0" q32="0.0" q33="0.0"/>
   <Multipole type="394" c0="1.613482" d1="0.0" d2="0.0" d3="0.0" q11="0.0" q21="0.0" q22="0.0" q31="0.0" q32="0.0" q33="0.0"/>
   <Multipole type="395" c0="0.252428" d1="0.0" d2="0.0" d3="0.0" q11="0.0" q21="0.0" q22="0.0" q31="0.0" q32="0.0" q33="0.0"/>
   <Multipole type="396" c0="2.048259" d1="0.0" d2="0.0" d3="0.0" q11="0.0" q21="0.0" q22="0.0" q31="0.0" q32="0.0" q33="0.0"/>
   <Multipole type="397" c0="-2.288234" d1="0.0" d2="0.0" d3="0.0" q11="0.0" q21="0.0" q22="0.0" q31="0.0" q32="0.0" q33="0.0"/>
   <Multipole type="398" c0="0.252428" d1="0.0" d2="0.0" d3="0.0" q11="0.0" q21="0.0" q22="0.0" q31="0.0" q32="0.0" q33="0.0"/>
   <Multipole type="399" c0="-1.581353" d1="0.0" d2="0.0" d3="0.0" q11="0.0" q21="0.0" q22="0.0" q31="0.0" q32="0.0" q33="0.0"/>
   <Polarize type="381" polarizability="5.048434386104e-04" thole="2.36164e-03" pgrp1="380" pgrp2="382" pgrp3="383" pgrp4="384" pgrp5="385" pgrp6="386" pgrp7="387" pgrp8="388" pgrp9="389" pgrp10="390" pgrp11="391" pgrp12="392" pgrp13="393" pgrp14="394" pgrp15="395" pgrp16="396" pgrp17="397" pgrp18="398" pgrp19="399"/>
   <Polarize type="382" polarizability="5.048434386104e-08" thole="2.36164e-03" pgrp1="380" pgrp2="381" pgrp3="383" pgrp4="384" pgrp5="385" pgrp6="386" pgrp7="387" pgrp8="388" pgrp9="389" pgrp10="390" pgrp11="391" pgrp12="392" pgrp13="393" pgrp14="394" pgrp15="395" pgrp16="396" pgrp17="397" pgrp18="398" pgrp19="399"/>
   <Polarize type="383" polarizability="5.048434386104e-08" thole="2.36164e-03" pgrp1="380" pgrp2="381" pgrp3="382" pgrp4="384" pgrp5="385" pgrp6="386" pgrp7="387" pgrp8="388" pgrp9="389" pgrp10="390" pgrp11="391" pgrp12="392" pgrp13="393" pgrp14="394" pgrp15="395" pgrp16="396" pgrp17="397" pgrp18="398" pgrp19="399"/>
   <Polarize type="384" polarizability="5.048434386104e-08" thole="2.36164e-03" pgrp1="380" pgrp2="381" pgrp3="382" pgrp4="383" pgrp5="385" pgrp6="386" pgrp7="387" pgrp8="388" pgrp9="389" pgrp10="390" pgrp11="391" pgrp12="392" pgrp13="393" pgrp14="394" pgrp15="395" pgrp16="396" pgrp17="397" pgrp18="398" pgrp19="399"/>
   <Polarize type="385" polarizability="5.048434386104e-08" thole="2.36164e-03" pgrp1="380" pgrp2="381" pgrp3="382" pgrp4="383" pgrp5="384" pgrp6="386" pgrp7="387" pgrp8="388" pgrp9="389" pgrp10="390" pgrp11="391" pgrp12="392" pgrp13="393" pgrp14="394" pgrp15="395" pgrp16="396" pgrp17="397" pgrp18="398" pgrp19="399"/>
   <Polarize type="386" polarizability="5.048434386104e-08" thole="2.36164e-03" pgrp1="380" pgrp2="381" pgrp3="382" pgrp4="383" pgrp5="384" pgrp6="385" pgrp7="387" pgrp8="388" pgrp9="389" pgrp10="390" pgrp11="391" pgrp12="392" pgrp13="393" pgrp14="394" pgrp15="395" pgrp16="396" pgrp17="397" pgrp18="398" pgrp19="399"/>
   <Polarize type="387" polarizability="5.048434386104e-08" thole="2.36164e-03" pgrp1="380" pgrp2="381" pgrp3="382" pgrp4="383" pgrp5="384" pgrp6="385" pgrp7="386" pgrp8="388" pgrp9="389" pgrp10="390" pgrp11="391" pgrp12="392" pgrp13="393" pgrp14="394" pgrp15="395" pgrp16="396" pgrp17="397" pgrp18="398" pgrp19="399"/>
   <Polarize type="388" polarizability="8.063631227791e-08" thole="2.36164e-03" pgrp1="380" pgrp2="381" pgrp3="382" pgrp4="383" pgrp5="384" pgrp6="385" pgrp7="386" pgrp8="387" pgrp9="389" pgrp10="390" pgrp11="391" pgrp12="392" pgrp13="393" pgrp14="394" pgrp15="395" pgrp16="396" pgrp17="397" pgrp18="398" pgrp19="399"/>
   <Polarize type="389" polarizability="8.063631227791e-08" thole="2.36164e-03" pgrp1="380" pgrp2="381" pgrp3="382" pgrp4="383" pgrp5="384" pgrp6="385" pgrp7="386" pgrp8="387" pgrp9="388" pgrp10="390" pgrp11="391" pgrp12="392" pgrp13="393" pgrp14="394" pgrp15="395" pgrp16="396" pgrp17="397" pgrp18="398" pgrp19="399"/>
   <Polarize type="390" polarizability="8.063631227791e-08" thole="2.36164e-03" pgrp1="380" pgrp2="381" pgrp3="382" pgrp4="383" pgrp5="384" pgrp6="385" pgrp7="386" pgrp8="387" pgrp9="388" pgrp10="389" pgrp11="391" pgrp12="392" pgrp13="393" pgrp14="394" pgrp15="395" pgrp16="396" pgrp17="397" pgrp18="398" pgrp19="399"/>
   <Polarize type="391" polarizability="8.063631227791e-08" thole="2.36164e-03" pgrp1="380" pgrp2="381" pgrp3="382" pgrp4="383" pgrp5="384" pgrp6="385" pgrp7="386" pgrp8="387" pgrp9="388" pgrp10="389" pgrp11="390" pgrp12="392" pgrp13="393" pgrp14="394" pgrp15="395" pgrp16="396" pgrp17="397" pgrp18="398" pgrp19="399"/>
   <Polarize type="392" polarizability="8.063631227791e-08" thole="2.36164e-03" pgrp1="380" pgrp2="381" pgrp3="382" pgrp4="383" pgrp5="384" pgrp6="385" pgrp7="386" pgrp8="387" pgrp9="388" pgrp10="389" pgrp11="390" pgrp12="391" pgrp13="393" pgrp14="394" pgrp15="395" pgrp16="396" pgrp17="397" pgrp18="398" pgrp19="399"/>
   <Polarize type="393" polarizability="8.063631227791e-08" thole="2.36164e-03" pgrp1="380" pgrp2="381" pgrp3="382" pgrp4="383" pgrp5="384" pgrp6="385" pgrp7="386" pgrp8="387" pgrp9="388" pgrp10="389" pgrp11="390" pgrp12="391" pgrp13="392" pgrp14="394" pgrp15="395" pgrp16="396" pgrp17="397" pgrp18="398" pgrp19="399"/>
   <Polarize type="394" polarizability="5.048434386104e-08" thole="2.36164e-03" pgrp1="380" pgrp2="381" pgrp3="382" pgrp4="383" pgrp5="384" pgrp6="385" pgrp7="386" pgrp8="387" pgrp9="388" pgrp10="389" pgrp11="390" pgrp12="391" pgrp13="392" pgrp14="393" pgrp15="395" pgrp16="396" pgrp17="397" pgrp18="398" pgrp19="399"/>
   <Polarize type="395" polarizability="5.048434386104e-08" thole="2.36164e-03" pgrp1="380" pgrp2="381" pgrp3="382" pgrp4="383" pgrp5="384" pgrp6="385" pgrp7="386" pgrp8="387" pgrp9="388" pgrp10="389" pgrp11="390" pgrp12="391" pgrp13="392" pgrp14="393" pgrp15="394" pgrp16="396" pgrp17="397" pgrp18="398" pgrp19="399"/>
   <Polarize type="396" polarizability="5.048434386104e-08" thole="2.36164e-03" pgrp1="380" pgrp2="381" pgrp3="382" pgrp4="383" pgrp5="384" pgrp6="385" pgrp7="386" pgrp8="387" pgrp9="388" pgrp10="389" pgrp11="390" pgrp12="391" pgrp13="392" pgrp14="393" pgrp15="394" pgrp16="395" pgrp17="397" pgrp18="398" pgrp19="399"/>
   <Polarize type="397" polarizability="5.048434386104e-08" thole="2.36164e-03" pgrp1="380" pgrp2="381" pgrp3="382" pgrp4="383" pgrp5="384" pgrp6="385" pgrp7="386" pgrp8="387" pgrp9="388" pgrp10="389" pgrp11="390" pgrp12="391" pgrp13="392" pgrp14="393" pgrp15="394" pgrp16="395" pgrp17="396" pgrp18="398" pgrp19="399"/>
   <Polarize type="398" polarizability="5.048434386104e-08" thole="2.36164e-03" pgrp1="380" pgrp2="381" pgrp3="382" pgrp4="383" pgrp5="384" pgrp6="385" pgrp7="386" pgrp8="387" pgrp9="388" pgrp10="389" pgrp11="390" pgrp12="391" pgrp13="392" pgrp14="393" pgrp15="394" pgrp16="395" pgrp17="396" pgrp18="397" pgrp19="399"/>
   <Polarize type="399" polarizability="5.048434386104e-08" thole="2.36164e-03" pgrp1="380" pgrp2="381" pgrp3="382" pgrp4="383" pgrp5="384" pgrp6="385" pgrp7="386" pgrp8="387" pgrp9="388" pgrp10="389" pgrp11="390" pgrp12="391" pgrp13="392" pgrp14="393" pgrp15="394" pgrp16="395" pgrp17="396" pgrp18="397" pgrp19="398"/>
 </AmoebaMultipoleForce>
  <AmoebaUreyBradleyForce cubic="0.0" quartic="0.0">
   <UreyBradley class1="74" class2="73" class3="74" k="-4.31294e+03" d="1.535676676685e-01"/>
 </AmoebaUreyBradleyForce>
</ForceField>
\end{lstlisting}

\newpage
\clearpage


\section{Error Reduction in MDCMs}
In the following it is demonstrated that applying constraints to
charge magnitudes during MDCM fitting can reduce the corresponding
error in electrostatic interaction energy between molecules of
arbitrary types and relative spatial orientations.

The MDCM optimal fitted solution without applied constraints minimizes
errors in the MEP, rather than the electrostatic interaction energy
relevant to MD simulations. Fitting to the electrostatic interaction
energy directly would avoid this issue, but also lead to biased
sampling of the electric field around the molecule during fitting by
focusing on those regions relevant to the electrostatic interactions
included in the finite training set. It is more difficult to generate
sufficiently large, heterogeneous training sets of reference
electrostatic interaction energies between non-polarized monomers than
to generate the more straightforward MEP across a spatial grid. A good
solution is therefore to solve the more straightforward MEP fitting
problem with constraints introduced to ensure that the resulting
charge models will also perform well for the closely related
electrostatic interaction energy between molecules.

To relate the error in the MEP to the resulting error in the
electrostatic interaction between molecules, we start with a
description of the residual error in the MEP of an MDCM for molecule
`a' after fitting:
\begin{equation}
V_{\rm a}({\rm \bf r}) = {\int_{\rm \Omega_a} { \frac{\rho_{\rm
        a}({\rm \bf r_a})}{|{\rm \bf r - r_{a}}|}}} dr_{\rm a} =
\sum_{n=1}^{n_{q,\rm a}} \frac{q_{{\rm a},n}}{|{\rm\bf r} - {\rm\bf
    r}_{{\rm\bf a},n}|} + \delta V_{ \rm a }( \rm\bf r ) \label{eq1}
\end{equation}
for ${\rm\bf r}$ sampled outside the molecular surface used for
fitting. $\rho_{\rm a}({\rm\bf r_{a}})$ is the reference electron
density of molecule `a' at point ${\rm\bf r_{a}}$ in the molecular
volume $\Omega_{a}$. $q_{{\rm a},n}$ is charge $n$ of the fitted MDCM
model with ${\rm\bf r}_{{\rm\bf a},n}$ the position of the charge, and
$\delta V_{ \rm a }( {\rm\bf r} )$ is the residual error in the MEP
after fitting the MDCM.

This can be related to the error in interaction with a single external
charge, $q_{\rm b}$ at ${\rm\bf r_{b}}$:

\begin{equation}
U_{ \rm ab} = q_{\rm b} \int_{\rm \Omega_{\rm a}} \frac{\rho_{\rm a}({\rm\bf r_a})}{|{\rm\bf r_{b} - r_{a}}|} dr_{\rm a}
 =  q_{\rm b} \sum_{n=1}^{n_{q,\rm a}} \frac{q_{{\rm a},n}}{|{{\rm\bf r_{b}} - {\rm\bf r}_{{\rm\bf a},n}}|} + q_{\rm b} \delta V_{ \rm a }({\rm\bf r_{b}})
\end{equation}

Using the exact interaction between two molecules, a double integral
over the molecular volumes of $a$ and $b$:
\begin{equation}
U_{\rm ab} = \int_{\rm \Omega_a} \int_{\rm \Omega_b} \frac{\rho_{\rm
    a}(\rm\bf r_a) \rho_{\rm b}(\rm\bf r_b)}{|{\rm\bf r_{b} - \rm\bf
    r_{a}}|} dr_{\rm a} dr_{ \rm b }
\end{equation}
and inserting \eqref{eq1} and rearranging yields
\begin{equation}
U_{\rm ab} = \sum_{n=1}^{\rm n_{q,a}} q_{{\rm a},n} \int_{\rm
  \Omega_b} \frac{\rho_{\rm b} (\rm\bf r_b)}{|{{\rm\bf r_{b}} -
    {\rm\bf r}_{{\rm\bf a},n}}|} + \int_{\Omega_{b}} \rho_{\rm
  b}({\rm\bf r_{b}}) \delta V_{\rm a}({\rm\bf r_{b}}) dr_{ \rm b }
\end{equation}
Using \eqref{eq1} again for molecule $b$:
\begin{equation}
U_{ \rm ab} = \sum_{n=1}^{\rm n_{q,a}} \sum_{m=1}^{\rm n_{q,b}}
\frac{q_{{\rm a},n} q_{{\rm b},m}}{|{{\rm\bf r}_{{\rm\bf b},m} -
    {\rm\bf r}_{{\rm\bf a},n}}|} + \sum_{n=1}^{n_{q,{\rm a}}} q_{{\rm
    a},n} \delta V_{\rm b}({\rm\bf r}_{{\rm\bf a},n}) +
\int_{\Omega_{b}} \rho_{\rm b}({\rm\bf r_{b}}) \delta V_{\rm
  a}({\rm\bf r_{b}}) dr_{ \rm b } \label{eq5}
\end{equation}
or, equivalently:
\begin{equation}
U_{ \rm ab} = \sum_{n=1}^{\rm n_{q,b}} \sum_{m=1}^{\rm n_{q,a}}
\frac{q_{{\rm b},n} q_{{\rm a},m}}{|{{\rm\bf r}_{{\rm\bf a},m} -
    {\rm\bf r}_{{\rm\bf b},n}}|} + \sum_{n=1}^{n_{q,{\rm b}}} q_{{\rm
    b},n} \delta V_{\rm a}({\rm\bf r}_{{\rm\bf b},n}) +
\int_{\Omega_{a}} \rho_{\rm a}({\rm\bf r_{a}}) \delta V_{\rm
  b}({\rm\bf r_{a}}) dr_{ \rm a } \label{eq6}
\end{equation}

\noindent
Minimizing the $\delta V_{\rm a}({\bf r})$ error term in \eqref{eq1}
requires simply a charge model that describes the MEP as closely as
possible to the corresponding Coulomb integral over the molecular
charge density. In practice this often results in large fitted
charges, of the order of several a.u. The error terms in \eqref{eq5}
and \eqref{eq6} for the electrostatic interaction energy, however,
show that the error in the MEP is multiplied by the magnitude of each
interacting charge. While error cancellation is possible it is not
guaranteed, so the presence of larger charges can lead to an
amplification of the error in the MEP. For a typical fitting problem
it has been observed that for a given MDCM model containing large
charges, the error in the electrostatic interaction energy is
amplified in some regions but remains small due to error cancellation
in others. It is therefore safest to constrain the magnitude of the
fitted charges to remain as low as possible while still achieving an
acceptable error in the fitted MEP, i.e. to find a compromise solution
with low errors in the fitted MEP and the smallest possible charges.

\newpage
\clearpage


\section{iAMOEBA MDCM OpenMM force field definition file}
\label{section:iAMOEBA-par}
The following "XML" format force field definition file defines the
iAMOEBA/MDCM implementation for use in OpenMM. Dummy atom types could
again potentially be consolidated as all MDCM dummy atoms also share
the same mass and VDW, and sites belonging to the same atom-type also
have the same polarizability.

\lstset{language=XML}
\begin{lstlisting}
<ForceField>
 <AtomTypes>
  <Type name="380" class="73" element="O" mass="15.999"/>
  <Type name="381" class="74" element="H" mass="1.008"/>
  <Type name="382" class="75" mass="0"/>
  <Type name="383" class="76" mass="0"/>
  <Type name="384" class="77" mass="0"/>
  <Type name="385" class="78" mass="0"/>
  <Type name="386" class="79" mass="0"/>
  <Type name="387" class="80" mass="0"/>
  <Type name="388" class="81" mass="0"/>
  <Type name="389" class="82" mass="0"/>
  <Type name="390" class="83" mass="0"/>
  <Type name="391" class="84" mass="0"/>
 </AtomTypes>
 <Residues>
  <Residue name="HOH">
   <Atom name="O" type="380"/>
   <Atom name="H1" type="381"/>
   <Atom name="H2" type="381"/>
   <Atom name="M1" type="382"/>
   <Atom name="M2" type="383"/>
   <Atom name="M3" type="384"/>
   <Atom name="M4" type="385"/>
   <Atom name="M5" type="386"/>
   <Atom name="M6" type="387"/>
   <Atom name="M7" type="388"/>
   <Atom name="M8" type="389"/>
   <Atom name="M9" type="390"/>
   <Atom name="M10" type="391"/>
   <Bond from="0" to="2"/>
   <Bond from="0" to="1"/>

<!-- H1-atom charges and positions -->
   <VirtualSite type="localCoords" index="3" atom1="0" atom2="1" atom3="2"  wo1="0.0" wo2="1.0" wo3="0.0"  wx1="-1.0" wx2="1.0"  wx3="0.0" wy1="-1.0" wy2="0.0" wy3="1.0"  p1=" 0.034734"  p2="-0.002091"  p3=" 0.000000" />
   <VirtualSite type="localCoords" index="4" atom1="0" atom2="1" atom3="2"  wo1="0.0" wo2="1.0" wo3="0.0"  wx1="-1.0" wx2="1.0"  wx3="0.0" wy1="-1.0" wy2="0.0" wy3="1.0"  p1="-0.016513"  p2=" 0.004524"  p3=" 0.000000" />
   <VirtualSite type="localCoords" index="5" atom1="0" atom2="1" atom3="2"  wo1="0.0" wo2="1.0" wo3="0.0"  wx1="-1.0" wx2="1.0"  wx3="0.0" wy1="-1.0" wy2="0.0" wy3="1.0"  p1="-0.005283"  p2=" 0.019053"  p3=" 0.000000" />
   <VirtualSite type="localCoords" index="6" atom1="0" atom2="1" atom3="2"  wo1="0.0" wo2="1.0" wo3="0.0"  wx1="-1.0" wx2="1.0"  wx3="0.0" wy1="-1.0" wy2="0.0" wy3="1.0"  p1="-0.005651"  p2="-0.014961"  p3="-0.000189" />
<!-- O-atom charges and positions -->
   <VirtualSite type="localCoords" index="7" atom1="0" atom2="1" atom3="2"  wo1="1.0" wo2="0.0" wo3="0.0"  wx1="-1.0" wx2="1.0"  wx3="0.0" wy1="-1.0" wy2="0.0" wy3="1.0"  p1="-0.002946"  p2="-0.010093"  p3=" 0.040674" />
   <VirtualSite type="localCoords" index="8" atom1="0" atom2="1" atom3="2"  wo1="1.0" wo2="0.0" wo3="0.0"  wx1="-1.0" wx2="1.0"  wx3="0.0" wy1="-1.0" wy2="0.0" wy3="1.0"  p1="-0.002031"  p2="-0.008820"  p3="-0.031699" />
<!-- H2-atom charges and positions -->
   <VirtualSite type="localCoords" index="9" atom1="0" atom2="1" atom3="2"  wo1="0.0" wo2="0.0" wo3="1.0"  wx1="-1.0" wx2="0.0"  wx3="1.0" wy1="1.0" wy2="-1.0" wy3="0.0"  p1=" 0.002756"  p2="-0.016593"  p3="-0.000082" />
   <VirtualSite type="localCoords" index="10" atom1="0" atom2="1" atom3="2"  wo1="0.0" wo2="0.0" wo3="1.0"  wx1="-1.0" wx2="0.0"  wx3="1.0" wy1="1.0" wy2="-1.0" wy3="0.0"  p1="-0.004779"  p2=" 0.018380"  p3="-0.000209" />
   <VirtualSite type="localCoords" index="11" atom1="0" atom2="1" atom3="2"  wo1="0.0" wo2="0.0" wo3="1.0"  wx1="-1.0" wx2="0.0"  wx3="1.0" wy1="1.0" wy2="-1.0" wy3="0.0"  p1="-0.037642"  p2="-0.005199"  p3="-0.000000" />
   <VirtualSite type="localCoords" index="12" atom1="0" atom2="1" atom3="2"  wo1="0.0" wo2="0.0" wo3="1.0"  wx1="-1.0" wx2="0.0"  wx3="1.0" wy1="1.0" wy2="-1.0" wy3="0.0"  p1=" 0.020099"  p2="-0.003550"  p3="-0.000000" />
 </Residue>
 </Residues>
  <AmoebaBondForce bond-cubic="-25.5" bond-quartic="379.3125">
   <Bond class1="73" class2="74" length="9.584047e-02" k="2.3331232e+05"/>
 </AmoebaBondForce>
  <AmoebaAngleForce angle-cubic="-0.014" angle-quartic="5.6e-05" angle-pentic="-7e-07" angle-sextic="2.2e-08">
   <Angle class1="74" class2="73" class3="74" k="6.359379296918e-02" angle1="1.064826e+02"/>
 </AmoebaAngleForce>
  <AmoebaOutOfPlaneBendForce type="ALLINGER" opbend-cubic="-0.014" opbend-quartic="5.6e-05" opbend-pentic="-7e-07" opbend-sextic="2.2e-08">
    <!-- LPW: Mark's force field parsing code requires AmoebaOutOfPlaneBendForce in order to read AmoebaAngleForce, even if the clause is empty -->
 </AmoebaOutOfPlaneBendForce>
  <AmoebaVdwForce type="BUFFERED-14-7" radiusrule="CUBIC-MEAN" radiustype="R-MIN" radiussize="DIAMETER" epsilonrule="HHG" vdw-13-scale="0.0" vdw-14-scale="1.0" vdw-15-scale="1.0">
   <Vdw class="73" sigma="3.645297e-01" epsilon="8.2348e-01" reduction="1.0"/>
   <Vdw class="74" sigma="0.0" epsilon="0.0" reduction="1.0"/>
   <Vdw class="75" sigma="0.0" epsilon="0.0" reduction="1.0"/>
   <Vdw class="76" sigma="0.0" epsilon="0.0" reduction="1.0"/>
   <Vdw class="77" sigma="0.0" epsilon="0.0" reduction="1.0"/>
   <Vdw class="78" sigma="0.0" epsilon="0.0" reduction="1.0"/>
   <Vdw class="79" sigma="0.0" epsilon="0.0" reduction="1.0"/>
   <Vdw class="80" sigma="0.0" epsilon="0.0" reduction="1.0"/>
   <Vdw class="81" sigma="0.0" epsilon="0.0" reduction="1.0"/>
   <Vdw class="82" sigma="0.0" epsilon="0.0" reduction="1.0"/>
   <Vdw class="83" sigma="0.0" epsilon="0.0" reduction="1.0"/>
   <Vdw class="84" sigma="0.0" epsilon="0.0" reduction="1.0"/>
 </AmoebaVdwForce>
   <AmoebaMultipoleForce direct11Scale="0.0" direct12Scale="1.0" direct13Scale="1.0" direct14Scale="1.0" mpole12Scale="0.0" mpole13Scale="0.0" mpole14Scale="0.4" mpole15Scale="0.8" mutual11Scale="1.0" mutual12Scale="1.0" mutual13Scale="1.0" mutual14Scale="1.0" polar12Scale="0.0" polar13Scale="0.0" polar14Intra="0.5" polar14Scale="1.0" polar15Scale="1.0">
   <Multipole type="380" c0="0.0" d1="0.0" d2="0.0" d3="0.0" q11="0.0" q21="0.0" q22="0.0" q31="0.0" q32="0.0" q33="0.0"/>
   <Multipole type="381" c0="0.0" d1="0.0" d2="0.0" d3="0.0" q11="0.0" q21="0.0" q22="0.0" q31="0.0" q32="0.0" q33="0.0"/>
   <Multipole type="382" c0="-0.106662" d1="0.0" d2="0.0" d3="0.0" q11="0.0" q21="0.0" q22="0.0" q31="0.0" q32="0.0" q33="0.0"/>
   <Multipole type="383" c0="-0.972581" d1="0.0" d2="0.0" d3="0.0" q11="0.0" q21="0.0" q22="0.0" q31="0.0" q32="0.0" q33="0.0"/>
   <Multipole type="384" c0="0.599708" d1="0.0" d2="0.0" d3="0.0" q11="0.0" q21="0.0" q22="0.0" q31="0.0" q32="0.0" q33="0.0"/>
   <Multipole type="385" c0="0.828910" d1="0.0" d2="0.0" d3="0.0" q11="0.0" q21="0.0" q22="0.0" q31="0.0" q32="0.0" q33="0.0"/>
   <Multipole type="386" c0="-0.236929" d1="0.0" d2="0.0" d3="0.0" q11="0.0" q21="0.0" q22="0.0" q31="0.0" q32="0.0" q33="0.0"/>
   <Multipole type="387" c0="-0.314238" d1="0.0" d2="0.0" d3="0.0" q11="0.0" q21="0.0" q22="0.0" q31="0.0" q32="0.0" q33="0.0"/>
   <Multipole type="388" c0="0.546572" d1="0.0" d2="0.0" d3="0.0" q11="0.0" q21="0.0" q22="0.0" q31="0.0" q32="0.0" q33="0.0"/>
   <Multipole type="389" c0="0.620478" d1="0.0" d2="0.0" d3="0.0" q11="0.0" q21="0.0" q22="0.0" q31="0.0" q32="0.0" q33="0.0"/>
   <Multipole type="390" c0="-0.624493" d1="0.0" d2="0.0" d3="0.0" q11="0.0" q21="0.0" q22="0.0" q31="0.0" q32="0.0" q33="0.0"/>
   <Multipole type="391" c0="-0.340765" d1="0.0" d2="0.0" d3="0.0" q11="0.0" q21="0.0" q22="0.0" q31="0.0" q32="0.0" q33="0.0"/>
   <Polarize type="380" polarizability="8.063631227791e-04" thole="2.36164e-03" pgrp1="381" pgrp2="382" pgrp3="383" pgrp4="384" pgrp5="385" pgrp6="386" pgrp7="387" pgrp8="388" pgrp9="389" pgrp10="390" pgrp11="391"/>
   <Polarize type="381" polarizability="5.048434386104e-04" thole="2.36164e-03" pgrp1="380" pgrp2="382" pgrp3="383" pgrp4="384" pgrp5="385" pgrp6="386" pgrp7="387" pgrp8="388" pgrp9="389" pgrp10="390" pgrp11="391"/>
   <Polarize type="382" polarizability="5.048434386104e-08" thole="2.36164e-03" pgrp1="380" pgrp2="381" pgrp3="383" pgrp4="384" pgrp5="385" pgrp6="386" pgrp7="387" pgrp8="388" pgrp9="389" pgrp10="390" pgrp11="391"/>
   <Polarize type="383" polarizability="5.048434386104e-08" thole="2.36164e-03" pgrp1="380" pgrp2="381" pgrp3="382" pgrp4="384" pgrp5="385" pgrp6="386" pgrp7="387" pgrp8="388" pgrp9="389" pgrp10="390" pgrp11="391"/>
   <Polarize type="384" polarizability="5.048434386104e-08" thole="2.36164e-03" pgrp1="380" pgrp2="381" pgrp3="382" pgrp4="383" pgrp5="385" pgrp6="386" pgrp7="387" pgrp8="388" pgrp9="389" pgrp10="390" pgrp11="391"/>
   <Polarize type="385" polarizability="5.048434386104e-08" thole="2.36164e-03" pgrp1="380" pgrp2="381" pgrp3="382" pgrp4="383" pgrp5="384" pgrp6="386" pgrp7="387" pgrp8="388" pgrp9="389" pgrp10="390" pgrp11="391"/>
   <Polarize type="386" polarizability="8.063631227791e-08" thole="2.36164e-03" pgrp1="380" pgrp2="381" pgrp3="382" pgrp4="383" pgrp5="384" pgrp6="385" pgrp7="387" pgrp8="388" pgrp9="389" pgrp10="390" pgrp11="391"/>
   <Polarize type="387" polarizability="8.063631227791e-08" thole="2.36164e-03" pgrp1="380" pgrp2="381" pgrp3="382" pgrp4="383" pgrp5="384" pgrp6="385" pgrp7="386" pgrp8="388" pgrp9="389" pgrp10="390" pgrp11="391"/>
   <Polarize type="388" polarizability="5.048434386104e-08" thole="2.36164e-03" pgrp1="380" pgrp2="381" pgrp3="382" pgrp4="383" pgrp5="384" pgrp6="385" pgrp7="386" pgrp8="387" pgrp9="389" pgrp10="390" pgrp11="391"/>
   <Polarize type="389" polarizability="5.048434386104e-08" thole="2.36164e-03" pgrp1="380" pgrp2="381" pgrp3="382" pgrp4="383" pgrp5="384" pgrp6="385" pgrp7="386" pgrp8="387" pgrp9="388" pgrp10="390" pgrp11="391"/>
   <Polarize type="390" polarizability="5.048434386104e-08" thole="2.36164e-03" pgrp1="380" pgrp2="381" pgrp3="382" pgrp4="383" pgrp5="384" pgrp6="385" pgrp7="386" pgrp8="387" pgrp9="388" pgrp10="399" pgrp11="391"/>
   <Polarize type="391" polarizability="5.048434386104e-08" thole="2.36164e-03" pgrp1="380" pgrp2="381" pgrp3="382" pgrp4="383" pgrp5="384" pgrp6="385" pgrp7="386" pgrp8="387" pgrp9="388" pgrp10="389" pgrp11="390"/>
 </AmoebaMultipoleForce>
  <AmoebaUreyBradleyForce cubic="0.0" quartic="0.0">
   <UreyBradley class1="74" class2="73" class3="74" k="-4.31294e+03" d="1.535676676685e-01"/>
 </AmoebaUreyBradleyForce>
 </ForceField>
\end{lstlisting}

\newpage
\clearpage


\section{Impact of Broken Symmetry in MDCMs}
In this section the H$_{2}$O dimer and cluster energies
are compared before and after rotating all even-numbered monomers
(i.e. the 2$^{\rm nd}$, 4$^{\rm th}$, 6$^{\rm th}$...) by 180 degrees
about their bisector in the PDB coordinate file. This effectively
amounts to exchanging atom indices of the hydrogen atoms and
quantifies the degree of symmetry-related differences in interaction
energies. As MDCMs employ no symmetry constraints during fitting in
the current work there will be a small discrepancy in interaction
energy after rotating monomers. This effect should be minimal if the
MDCM is sufficiently well fitted to the symmetric MEP around the
molecule, but is non-existent in the multipolar and DCM models that
share the spatial symmetry of the equilibrium structure of the molecule
itself. The magnitude of the discrepancy is shown in Table
\ref{tab:si-mdcm-asymm}. It should also be pointed out that even
  if the equilibrium structure of a molecule has a particular
  point-group symmetry and the distribution of charges shares this
  symmetry (whether for conventional PCs or for (M)DCM models), during
  an MD simulation the instantaneous geometries typically have lower
  symmetry. 

\begin{table}
\caption{Comparison between multipolar and MDCM iAMOEBA electrostatic
  energies, including polarization. MDCM energies are calculated with
  all H$_{2}$O monomers in their original orientations (orig), and
  with even-numbered monomers rotated 180 degrees about their bisector
  from their original orientations (rotated / rot) to evaluate the
  impact of symmetry violations in the MDCMs on cluster electrostatic
  energies. Energy differences are presented between MDCM monomers in
  their original orientations and their multipolar iAMOEBA equivalents
  ($\Delta E_{\rm orig-mtp}$), the same comparison is made using
  rotated monomers and multipolar iAMOEBA ($\Delta E_{\rm rot-mtp}$),
  and finally the energies of the original MDCM orientations are
  directly compared to MDCM energies with rotated monomers ($\Delta
  E_{\rm orig-rot}$). In the top part of the table energies for the 10
  dimer structures are shown, while in the bottom part energies for
  various oligomers are presented.}
\begin{tabular}{l |c*{7}{c}r}
Dimer  &iAMOEBA   &Original& $\Delta E_{\rm orig-mtp}$ & Rotated & $\Delta E_{\rm rot-mtp}$ & $\Delta E_{\rm orig-rot}$\\
       &(kcal/mol)&(kcal/mol)  & (kcal/mol) & (kcal/mol)& (kcal/mol)& (kcal/mol)&\\
\hline
1 & --5.113 & --5.125 & --0.011 & --5.110 & 0.003 & 0.014\\
2 & --4.505 & --4.465 & 0.040 & --4.491 & 0.013 & --0.027\\
3 & --4.504 & --4.471 & 0.033 & --4.467 & 0.037 & 0.004\\
4 & --3.824 & --3.836 & --0.011 & --3.814 & 0.011 & 0.022\\
5 & --3.269 & --3.180 & 0.089 & --3.274 & --0.005 & --0.011\\
6 & --2.972 & --2.863 & 0.109 & --2.990 & --0.018 & 0.020\\
7 & --3.199 & --3.196 & 0.003 & --3.195 & 0.004 & 0.001\\
8 & --1.572 & --1.587 & --0.014 & --1.577 & --0.005 & 0.009\\
9 & --3.794 & --3.770 & 0.024 & --3.770 & 0.025 & 0.000\\
10 & --3.016 & --3.026 & --0.010 & --3.025 & --0.009 & 0.001\\
\hline
Oligomer  &iAMOEBA   &Original& $\Delta E_{\rm orig-mtp}$ & Rotated & $\Delta E_{\rm rot-mtp}$ & $\Delta E_{\rm orig-rot}$\\
       &(kcal/mol)&(kcal/mol)  & (kcal/mol) & (kcal/mol)& (kcal/mol)& (kcal/mol)&\\
\hline
trimer	&	--13.770	&	--13.782	&	--0.012	& --13.725 & 0.045 & 0.056\\
tetramer	&	--24.529	&	--24.436	&	0.093	& --24.457 & 0.071 & --0.022\\
pentamer	&	--32.321	&	--32.266	&	0.055	& --32.180 & 0.141 & 0.086\\
hexamer 	&	--41.459	&	--41.348	&	0.111	& --41.262 & 0.197 & 0.085\\
heptamer	&	--51.299	&	--51.129	&	0.171	& --51.009 & 0.290 & 0.119\\
octamer	&	--64.672	&	--64.238	&	0.434	& --64.166 & 0.506 & 0.072\\
nonamer	&	--72.896	&	--72.478	&	0.419	& --72.424 & 0.472 & 0.054\\
decamer	&	--82.858	&	--82.402	&	0.456	& --82.313 & 0.544 & 0.089\\
\label{tab:si-mdcm-asymm}
\end{tabular}
\end{table}

The difference in H$_{2}$O cluster electrostatic energies (including
polarization) before and after rotating half of the monomers ($\Delta
E_{\rm orig-rot}$) remains of the order of a few hundredths of a
kcal/mol for all clusters and does not increase significantly with
increasing cluster size. Together with results of bulk simulations,
where all monomers are free to rotate their relative orientations during
each trajectory, these results demonstrate that the spatial asymmetry
in the MDCM charge arrangement does not have a significant impact on
interaction energies or bulk simulation properties.\\

\noindent
While development of symmetry constraints would eliminate this effect
entirely, the fitting process would then be complicated. For
example, charges may enter or leave the symmetry planes of the
molecule during fitting. Each charge that leaves a plane must spawn an additional
charge on the opposite side of that plane to maintain symmetry, favoring
solutions with fewer in-plane charges as additional charges tend to
improve the fit, and complicating control over the total number of
charges in the fit. While modifications can be envisaged to deal with
such issues, the fitting process is likely to become more complex as a
result. As such modifications are additionally applicable to a
relatively small subset of available molecules (those with high symmetry) they are currently left as a
future development.\\

\noindent
For the purpose of the current study, namely demonstrating the
accuracy and performance of new developments to MDCMs with respect to
existing multipolar and TIP$n$P force field models, the results in
Table \ref{tab:si-mdcm-asymm} and the results section of the main text
show that symmetry constraints are not required. 

\newpage
\clearpage


\section{O...O Radial Distribution Functions}

\begin{figure}[!ht]
\centering
\begin{minipage}[b]{0.9\textwidth}
\includegraphics[width=0.9\textwidth]{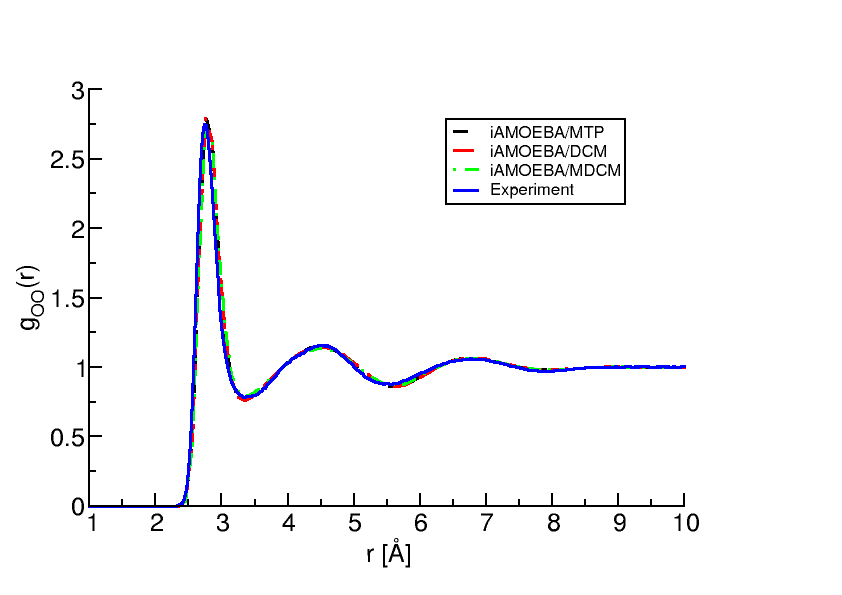}
\end{minipage}
\caption{Oxygen--oxygen radial distribution functions for liquid water
  at 298.15 K and 1 atm. DCM and MDCM implementations are compared
  with results using the original, multipolar water iAMOEBA water
  model and with experimental neutron diffraction
  data\cite{soper2000radial}.}
\end{figure}

\newpage
\clearpage


\section{O...H Radial Distribution Functions}

\begin{figure}[!ht]
\begin{center}
\includegraphics[height=12cm,angle=-90.0]{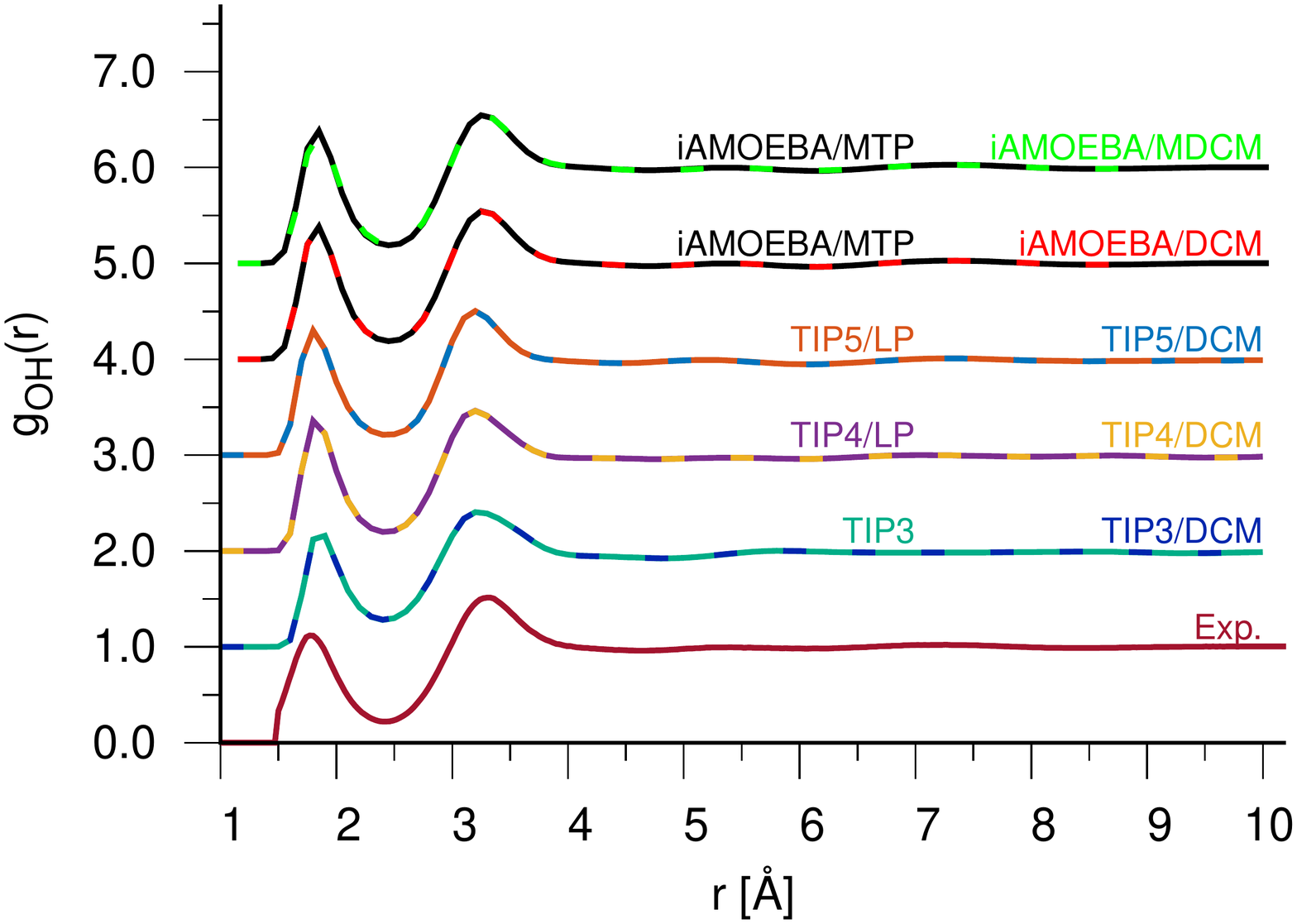}
\caption{Oxygen--hydrogen radial distribution functions for liquid
  water at 298.15 K and 1 atm. Various simulation methods are compared
  against experimental Neutron diffraction data
  \cite{soper2000radial}.  Successive curves are offset 1 unit along
  the $y$-axis for clarity.}
\label{fig:groh}
\end{center}
\end{figure}

\newpage
\clearpage


\section{H...H Radial Distribution Functions}

\begin{figure}[!ht]
\begin{center}
\includegraphics[height=10cm,angle=-90.0]{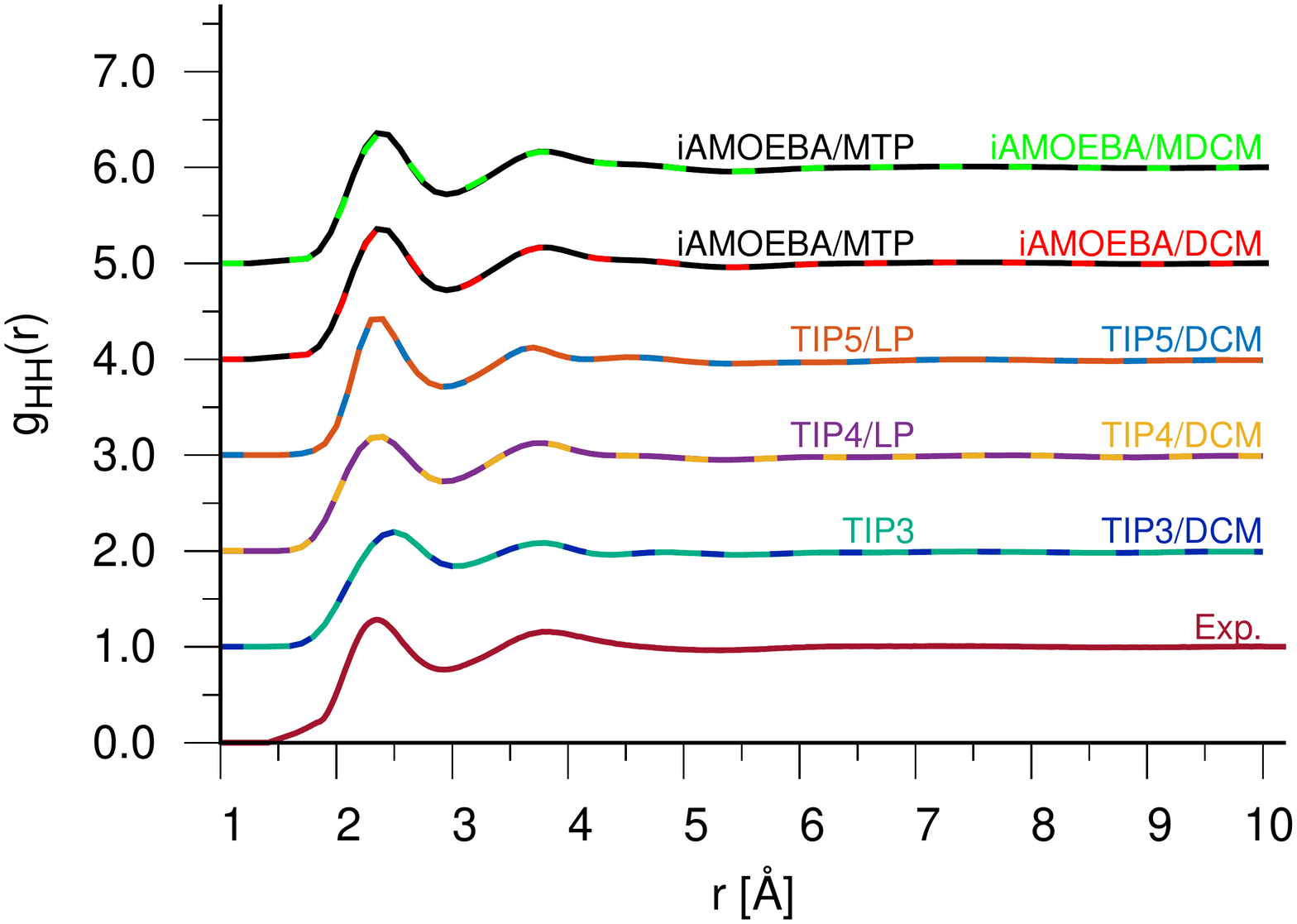}
\caption{Hydrogen--hydrogen, radial distribution functions for liquid
  water at 298.15 K and 1 atm. Various simulation methods are compared
  against experimental Neutron diffraction data
  \cite{soper2000radial}.  Successive curves are offset 1 unit along
  the $y$-axis for clarity.}
\label{fig:grhh}
\end{center}
\end{figure}

\newpage
\clearpage


\section{Computational Efficiency}
As indicated in section \ref{sec:reduction}, using (M)DCM
over MTPs may also have a beneficial effect on computational
efficiency in MD simulations. It is first noted that both approaches
scale fundamentally as $O(N)$ or $O(NlogN)$ where $N$ is the number of
atoms\cite{nilsson2009} as the number of interacting atoms remains the
same in both cases, and the multiple charges or multipole terms
belonging to each atom do not interact with one another. The relative
number of terms for MDCM compared with MTP for a basic implementation
using a spherical tensor formalism for the multipole expansion can be
gleaned from Equations \ref{Eq:v0000}--\ref{Eq:v22c22c}.

The cost for both, distributed charges and multipole moments,
therefore arises from the increased number of terms per atom--atom
interaction. For $n$ non-zero multipole terms and $n$ distributed
charges each atom--atom interaction involves $n(n+1)/2$ component
terms. As multipole models allow for truncation at different ranks for
different atoms,\cite{MM.mtp:2012} and MDCMs allow for different
numbers of charges per atom the general summation over terms per
atom--atom interaction is
\begin{equation}
\label{eq:pot-ab}
 V_{\rm ab}({\rm \bf r}_{\rm ab}) = \sum_{i=1}^{n_{\rm a}}\sum_{j=1}^{n_{\rm b}}V_{ij}({\rm \bf r}_{\rm ab})
\end{equation}
where $n_{\rm a}$ is the number of nonzero multipole moment components
or distributed charges on atom $a$, respectively. $V_{ij}({\rm \bf
  r}_{\rm ab})$ is the contribution to the interaction energy $V_{\rm
  ab}({\rm \bf r}_{\rm ab})$ between atoms $a$ and $b$ from multipole
moment components or charges $i$ and $j$ as a function of atomic
separation ${\rm \bf r}_{\rm ab}$ for both models, including the
relative orientations of the local axis systems for multipolar
models. Computational cost is therefore fundamentally linked to
$n_{\rm a}$, $n_{\rm b}$ and the functional form of $V_{ij}({\rm \bf
  r}_{\rm ab})$.\\

\noindent
Hence, substantial computational savings can be achieved by reducing
the number of terms $n_{\rm a}$ and $n_{\rm b}$, see Equation
\ref{eq:pot-ab}. For multipole moments this is typically achieved by
diagonalizing the Cartesian quadrupole moment tensor and converting to
spherical polar coordinates, leaving maximally 6 non-zero multipole
moment components. In highly symmetric systems this number may reduce
further,\cite{MM.mtp:2012} and the multipole expansion may be
truncated at lower rank for certain atom types such as hydrogen
atoms. In MDCMs the fitting algorithm is explicitly conceived to use
as few charges per atom as possible. In the current study it was
demonstrated that no more than 10 charges are required in total
compared to 19 non-zero multipole moment components in the iAMOEBA
model. For PhF, 22 charges were used for the entire molecule (see
Figure \ref{fig:phf-mdcm-fit} of the main text) compared with 46 non-zero multipole
components.\\

\noindent
Next, the cost of evaluating the coordinate-dependent terms
$V_{ij}({\rm \bf r}_{\rm ab})$ is considered. A comparison of these
terms has been presented previously\cite{devereux2014novel} but is
nevertheless discussed in some detail here.
\begin{itemize}

\item
While the use of off-centered charges introduces complexity in
evaluation of local axes to place the charges, these terms are the
same for all atom--atom interactions and are evaluated once per atom
per simulation time step. For multipoles the local axes also need to
be evaluated to determine the torques.

\item
Conversely, each Coulomb charge--charge term or multipole--multipole
term and its derivative must be evaluated for every atom--atom
interaction and so dominates the computational cost. For multipole
moments higher ranking terms incur increasing computational cost due
to the need to evaluate trigonometric functions (Equations
\ref{Eq:v0000}--\ref{Eq:v22c22c}), although schemes exist to reduce
this.\cite{brooksMTP} For MDCM there is a small increase in complexity
of the forces with respect to an atom-centered charge model to include
the axis derivatives that are calculated once per time step, but as
shown in Equation \ref{eq:pot-ab} the form of the interaction energy
is the same as for an atom-centered charge model.

\item
For MDCM an additional inverse square root operation to evaluate $1/r$
for each off-nuclear charge position is incurred, whereas all
multipole moments share the same local origin for a given atom and
thus share the same value of $1/r$.
\end{itemize}

\noindent
The comparison of computational cost of $V_{ij}({\rm \bf r}_{\rm ab})$
is thus primarily a comparison of the cost of the additional inverse
square root operation per interaction $ij$ (MDCM), with the
significantly increased number of multiplication and division
operations and the trigonometric terms required especially for higher
order multipole terms. Based on this argument it also becomes clear
that in order to increase the accuracy of a multipolar model beyond
rank $l=2$ (quadrupole moment) a significant cost will be incurred due
to the large increase in $n_{\rm a}$ and $n_{\rm b}$ and the
additional increase in complexity of $V_{ij}({\rm \bf r}_{\rm ab})$,
whereas additional charges can be added one by one in MDCM with no
increase in complexity of $V_{ij}({\rm \bf r}_{\rm ab})$.\\

\noindent
The following timings are illustrative examples for two particular
implementations for two different systems used in the current work and
caution should be used before drawing general conclusions (see further
below). The first system studied was the iAMOEBA waterbox introduced
in the main text. To make a meaningful comparison the polarization
term was removed, as current OpenMM functionality required 13
multipolar, polarizable sites to be created for MDCM where only 3
polarizable sites (at the nuclear positions) and no multipole moments
would have been necessary with a minor modification to the code (see
also Section \ref{Section:MDCMinOpenMM} of the main text). The
polarization energy therefore dominated the calculation time and
hindered analysis of the cost of the underlying charge model. To
remove polarization the Multipole definitions of the OpenMM parameter
file (section \ref{section:iAMOEBA-par}) were replace by
``NonbondedForce" definitions of the form:

\lstset{language=XML}
\begin{lstlisting}
   <Atom name="M1" type="382" charge="-0.106662" sigma="0.0" epsilon="0.0"/>
\end{lstlisting}

To keep results comparable, polarizabilities were also zeroed when
using multipoles, and $10^5$ time steps (50 ps) equilibration was run
with PME for each model (see section
\ref{Section:MDCMinOpenMM}--\ref{Section:MDSimAndPropComp} of the main
text for simulation details). The initial potential energies for the
simulation water box without polarization are within 0.3 kcal/mol for
both electrostatic models (--688.5 kcal/mol for MDCM vs. --688.8
kcal/mol for MTP). The simulation wall clock time on a single NVidia
Titan Xp GPU with CUDA 10.2 and Intel(R) Xeon(R) E5-2620 v4 CPU was
97.8 s for MDCM (averaged over 5 runs) and 112.5 s for MTPs, i.e. a
speedup of $\sim 15$ \% for MDCM compared with MTP. It should be
emphasized that the code in OpenMM is not likely to be optimized for
MDCM, as models with more than one or two off-centered charges are
currently uncommon, and even the non-polarizable OpenMM MDCM
implementation leaves 10 superfluous van der Waals sites (1 per
off-center charge) that are not required by MDCM, and 3 superfluous
charged sites at the nuclear positions, while more focus on the
implementation of virtual sites could potentially lead to further
gains. In the same way the iAMOEBA multipolar implementation with
zeroed polarizabilities would be made more efficient by completely
removing the polarizable sites, which appear to incur a computational
cost even when polarizabilities are zeroed. Despite these caveats it
is encouraging that the MDCM representation runs efficiently, even
without the minor modifications to the simulation code suggested.\\

\noindent
A similar comparison was run for bulk PhF with spherical cut-offs in
CHARMM for 4000 time steps (4 ps) heating and 4000 time steps (4 ps)
equilibration of the pure liquid PhF simulation box described in
section \ref{section:PhFProps} of the main text with 8 CPU cores of an
Intel Xeon E5-2630 v4 CPU. In this case neither code has been heavily
optimized, but the average wall clock time averaged over 5 simulations
was 848 s (MDCM) vs. 3233 s (MTP). Again, it is gratifying to note the
decreased computational cost associated with the MDCM, but again both
timings describe one possible implementation of each algorithm in one
particular hardware and software environment for the PhF system
studied here, and caution should still be used before drawing general
conclusions.

The discussion to this point neglects details of the algorithms,
implementation, hardware, compilers and others that can make a
decisive difference to the efficiency for large scale
applications. Control and consideration of all of these points for
both approaches will be required for further quantitative comparison
of computational cost, and are beyond the scope of the current
analysis.
\newpage
\clearpage


\bibliography{si}